\documentclass[12pt,preprint]{aastex}

\begin{document}

\shorttitle{M Dwarf Metallicities}
\shortauthors{Bean et al.}

\title{Accurate M Dwarf Metallicities from Spectral Synthesis: A Critical Test of Model Atmospheres}

\author{Jacob L. Bean}
\affil{Dept.\ of Astronomy, University of Texas, 1 University Station, C1402, Austin, TX 78712}
\email{bean@astro.as.utexas.edu}

\author{Christopher Sneden}
\affil{Dept.\ of Astronomy, University of Texas, 1 University Station, C1402, Austin, TX 78712}
\email{chris@astro.as.utexas.edu}

\author{Peter H. Hauschildt}
\affil{Hamburger Sternwarte, Gojenbergsweg 112, 21029 Hamburg, Germany}
\email{yeti@hs.uni-hamburg.de}

\author{Christopher M. Johns-Krull}
\affil{Department of Physics \& Astronomy, Rice University, 6100 Main Street, Houston, TX 77005}
\email{cmj@rice.edu}

\and
\author{G. Fritz Benedict}
\affil{McDonald Observatory, University of Texas, 1 University Station, C1402, Austin, TX 78712}
\email{fritz@astro.as.utexas.edu}

\received{}
\revised{}
\accepted{}

\begin{abstract}
We describe a method for accurately determining M dwarf metallicities with spectral synthesis based on abundance analyses of visual binary stars. We obtained high resolution, high signal-to-noise spectra of each component of five visual binary pairs at McDonald Observatory. The spectral types of the components range from F7 to K3 V for the primaries and M0.5 to M3.5 V for the secondaries. We have determined the metallicities of the primaries differentially with respect to the Sun by fitting synthetic spectra to Fe \textsc{i} line profiles in the observed spectra. In the course of our analysis of the M dwarf secondaries we have made significant improvements to the PHOENIX cool-star model atmospheres and the spectrum analysis code MOOG. Our analysis yields a RMS deviation of 0.11 dex in metallicity values between the binary pairs. We estimate the uncertainties in the derived stellar parameters for the M dwarfs to be 48 K, 0.10 dex, 0.12 dex, 0.15 km s$^{-1}$, and 0.20 km s$^{-1}$ for {T}$_{eff}$, log \textsl{g}, [M/H], $\xi$, and $\eta$  respectively. Accurate stellar evolutionary models are needed to progress further in the analysis of cool-star spectra; the new model atmospheres warrant recalculation of the evolutionary models.

\end{abstract}

\keywords{stars: abundances -- stars: atmospheres -- stars: late-type -- binaries: visual -- stars: individual (HIP 12114, HIP 26907, HIP 32423, HIP 40035, HIP 102040)}

\section{INTRODUCTION}
The lowest mass stars, M dwarfs, are the most abundant stellar objects in our galaxy. They make up over 70\% of stars in number and contribute over 40\% of the total stellar mass content \citep{henry98}. Despite their numbers, these stars remain one of the least understood stellar types. This is due to a lack of empirical data needed to test the predictions of theoretical models. Metallicity, one such parameter for which precise data do not yet exist for this stellar type, is the motivation for this study. 

Metallicity, [M/H]\footnote{We adopt the standard spectroscopic notation: for elements X and Y, log $\epsilon$(X) $\equiv$ log$_{10}(N_{X}/N_{H})$ + 12.0, [X/Y] $\equiv$ log$_{10}(N_{X}/N_{Y})_{\star}$ - log$_{10}(N_{X}/N_{Y})_{\sun}$, and $N_{X}$ is the number density of element X. We assume [M/H] = [Fe/H] throughout this paper.}, is of particular importance in theoretical models of low-mass stars because of its role, along with mass, in governing early evolution and main sequence properties \citep{kroupa97, baraffe98, siess00}. Additionally, stellar metallicity has been shown to correlate with the occurrence of extrasolar giant planets \citep[most recently][]{fv05, santos05}. As planets are being discovered and confirmed around M dwarfs \citep{delfosse98,marcy98,marcy01,benedict02,rivera05,butler04,bonfils05}, the question of these stars' metallicities has become increasingly relevant.

The historic lack of data for M dwarfs has been due primarily to their intrinsic faintness, a consequence of their low mass. However, despite the proliferation of large aperture telescopes and increasingly sensitive instrumentation, precise determinations of the fundamental parameters for M dwarfs are still rare. The factor now limiting the determination of accurate metallicities is the lack of a technique reliable enough to interpret the complex spectra of these stars.

Modern analysis techniques applied to high-resolution, R, and high signal-to-noise, S/N, spectra of solar-type stars consistently yield chemical abundances with internal precisions of 10\% \citep[e.g.][]{carlos04,vf05}. The application of these methods to M dwarfs is complicated by the effects of significant molecule formation and the resulting opacity in the photospheres of these stars. Molecular band spectra are much more complex than atomic spectra and typically dominate the spectral regions in which they are located. The regions traditionally utilized to derive metallicities for solar-type stars, the visible and red, are affected by TiO absorption bands in M dwarfs. These TiO lines blend with all other lines and create a ``pseudo continuum,'' making equivalent width measurements of atomic lines in this region unreliable for all but the earliest M dwarfs \citep{woolf06}. Therefore, comprehensive synthetic spectrum analyses must be employed.

\citet[hereafter V98]{v98} pioneered the use of spectral synthesis to determine M dwarf parameters to high precision. They fit synthetic spectra of TiO and atomic lines based on the NextGen version of the PHOENIX model atmospheres \citep{haus99} to a high quality observed spectrum of the M3.5 V star Gl 725B to determine its parameters. Their precisions for the parameters effective temperatures, \textsl{T}$_{eff}$, surface gravity, log \textsl{g}, metallicity, [M/H], and macroturbulent velocity, $\eta$, were 71 K, 0.14 dex, 0.07 dex, and 0.7 km s$^{-1}$ respectively. 

In this paper we present the results of a test of the V98 technique and PHOENIX cool-star model atmospheres, and our work to improve both based on the abundance analyses of visual binary stars. In the next section we describe our high resolution, high S/N observations of five visual binary pairs containing primaries ranging from spectral type F7 V -– K3 V and secondaries M0.5 V -– M3.5 V. In \S 3 we discuss our differential abundance analysis of the primaries with respect to the Sun. In \S 4 we describe our application of the V98 technique to derive the metallicities of the M dwarf secondaries. Based on the assumption that stars in a bound system have the same metallicities, we show that the original V98 technique yields metallicities for M dwarfs that are systematically 0.5 dex too low. We then outline our updates and modifications to this technique in \S 5. We show that a re-analysis of the M dwarfs in our binary sample, using our revised method and new model atmospheres, gives a RMS deviation of 0.11 dex from the primaries' metallicities. In \S 6 we present an error analysis of the new technique, which yields uncertainties of 48 K for \textsl{T}$_{eff}$, 0.10 dex for log \textsl{g}, 0.12 dex for [M/H], 0.15 km s$^{-1}$ for the microturbulent velocity, $\xi$, and 0.20 km s$^{-1}$ for $\eta$. Finally, in \S 7, we discuss the potential ramifications of our validation of the PHOENIX cool-star model atmospheres and our future applications of this technique to large samples of M dwarfs.

\section{OBSERVATIONS AND DATA REDUCTION}
We selected five nearby, common proper motion, visual binary pairs containing a solar-similar primary and an M dwarf secondary from the \citet{poveda94}, \citet{allen00}, and \textsl{Hipparcos} \citep{hipp} catalogs. We use the term ``solar-similar'' here to refer to stars for which abundances can be derived spectroscopically in an identical manner as for the Sun, roughly main sequence mid-F to mid-K spectral class stars. Basic data for the selected pairs are listed in Table ~\ref{tab:table1}. Spectral types of the primaries span the range from F7 V -- K3 V. The M dwarf secondaries range from spectral type M0.5 V -- M3.5 V, with one object not having spectral type but assumed to be an M dwarf based on its placement in an HR diagram. The minimum component separation on the sky, $\rho$, for the five pairs is 31\arcsec. Therefore, each component was easily isolated and there was no light contamination from its companion.

We observed each component of the five visual binary pairs using the 2.7m Harlan J. Smith telescope at McDonald Observatory on November 20 and 21, 2003. Data were obtained with the 2dcoud\'{e} spectrograph \citep{tull95} equipped with a 79 gr mm$^{-1}$ echelle grating and 8.2\arcsec x 1.2\arcsec\ slit. Exposure times varied from 5 to 30 minutes. Multiple 30 minute exposures were taken for the M dwarfs and co-added to facilitate cosmic ray subtraction. The maximum total exposure time for a single object was 180 minutes. Additionally, we recorded a spectrum of the day sky via a port that directs outside light on to the slit entrance of the instrument. 

CCD reduction and optimal order extraction were carried out using the standard IRAF\footnote{IRAF is distributed by the National Optical Astronomy Observatories, which are operated by the Association of Universities for Research in Astronomy, Inc., under cooperative agreement with the National Science Foundation.} routines in the \textsl{imred}, \textsl{ccdred}, and \textsl{echelle} packages. After extraction, the spectra in each order were flattened using the IRAF task \textsl{continuum}. The wavelength calibrations for each night were calculated based on the identification of roughly 1000 lines in thorium-argon emission spectra taken at the beginning of the night and have RMS precisions of 0.002 \AA. Each exposure contains 37 echelle orders with incomplete coverage spanning the range 3800 -- 9900 \AA. The average order width is 110 \AA. Gaps between the orders begin at 4000 \AA\ and increase in size with wavelength. 

The final one-dimensional spectra have measured continuum S/N, assuming Poisson statistics, ranging from 245 -- 592 pixel$^{-1}$ for the primaries and 145 -- 300 pixel$^{-1}$ for the secondaries in the spectral regions used for analysis. The measured resolving power was roughly 50,000.

\section{ANALYSIS OF THE PRIMARIES}
We determined the metallicities of the primaries in our sample by using a technique that is a variation of the traditional approach for solar-similar stars. We constrain the stellar effective temperatures and surface gravities by using a photometric color relationship and evolutionary models respectively. We derive an iron abundance relative to the solar value by fitting synthetic spectra to the observed line profiles of Fe \textsc{i} lines and equate this parameter to metallicity. The details of our approach are discussed in the following subsections.

\subsection{Line Data}
\begin{figure}
\epsscale{1.0}
\plotone{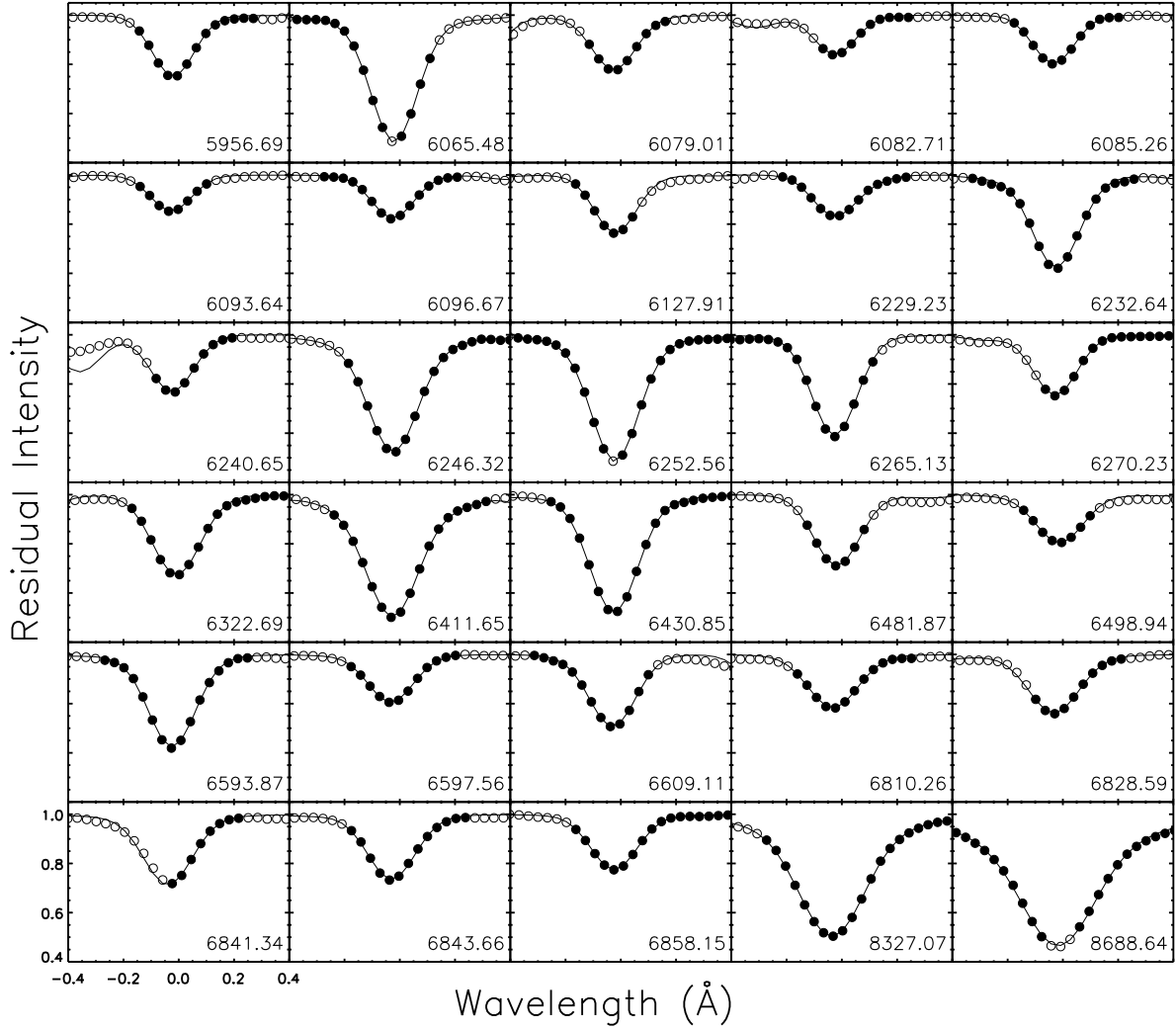}
\caption{Best individual fits of synthetic spectra (solid line) to the 30 Fe \textsc{i} line profiles (points) of the solar-similar primary HIP 102040A. Line center wavelengths are given in the corresponding panel. The filled points were used in the fitting process; the open points were ignored. The number of points ignored in the cores of the strong lines depends on their strength and therefore varies with the stellar parameters.}
\label{fig:f1}
\end{figure}

We began by identifying Fe \textsc{i} lines that are unblended with neighboring lines in our observed day-sky spectrum, which we considered a proxy for the solar spectrum. We limited our search to lines for which accurate laboratory data was available in the compilation of \citet{ramirez06} and van der Waals damping data had been calculated by \citet{barklem00}. For the purpose of continuum normalization, we required that each line have a region that was apparently free of contaminating lines within 3 \AA\ of the line center. We then examined the selected lines in the spectra of the primaries for blends and continuum windows to ensure their utility for these stars. In order to retain a reasonable sample of lines, we had to relax the unblended constraint. We identified portions of each line that did remain relatively free of contamination and used this information to construct a mask of spectral regions to be used in our analysis. Thirty Fe \textsc{i} lines were selected and their profiles in the observed spectrum of one of the primaries, HIP 102040A, are shown in Figure ~\ref{fig:f1}. Also shown is the fit, found with the procedure described below, used to determine the stellar parameters. Figure ~\ref{fig:f1} illustrates the relatively clean nature of the lines selected and the specific regions that we used in our analysis.

We then determined astrophysical log \textsl{gf}s for the selected lines. Our procedure was the inverse of the procedure we used to determine the iron abundances of the primaries and was a two step process. We adopted a model atmosphere with the standard solar parameters, \textsl{T}$_{eff}$ = 5777 K, log \textsl{g} = 4.44, and [M/H] = 0.0 (by definition) that was interpolated from the grid described in the following subsection. We assumed the solar abundances of \citet{asplund05}, namely log $\epsilon$(Fe)$_{\sun}$ = 7.45. Microturbulence and macroturbulence are not purely physical parameters and their adopted solar values vary greatly in the literature. Therefore, our first step in the process of determining astrophysical log \textsl{gf}s was to determine the solar microturbulence and macroturbulence values to be used. To do this we fit synthetic spectra to the high quality solar spectrum of \citet{kurucz84}. We determined the microturbulence, $\xi$, and Gaussian macroturbulence, $\eta$, values that yielded synthetic spectra that best reproduced the selected line profiles as a group using an adaptation of the $\chi^{2}$ minimization algorithm of Marquardt \citep{mar63, press86}. We assumed the \citet{ramirez06} lab log \textsl{gf} values for this step and found the solar $\xi$ to be 1.24 km s$^{-1}$ and $\eta$ to be 1.90 km s$^{-1}$.

With these determined line broadening parameters and our adopted solar model atmosphere and abundances, we then determined log \textsl{gf}s values that best reproduced the solar line profiles. We again used the $\chi^{2}$ minimization algorithm to find which values gave synthetic spectra that best fit the Kurucz solar spectrum. The final line data, including the lab log \textsl{gf}s for comparison, are listed in Table ~\ref{tab:table2}. The astrophysical log \textsl{gf} values average 0.08 dex lower than the lab values. As shown in \S 3.4, our systematic adjustment of the line data is inconsequential because our analysis of the primaries is purely differential to the Sun.

\subsection{Model Atmospheres}
We chose to use model atmospheres computed with PHOENIX for our analysis of the primaries in order to maintain consistency with our analysis of the M dwarf secondaries. The models are the latest version of the NextGen version models presented by \citet{haus99} and are discussed further in \S 5.2. We generated a grid of models with version 13 of PHOENIX for the analysis of the primaries spanning the ranges 4000 $\le$ \textsl{T}$_{eff}$ $\le$ 7000 K, 3.5 $\le$ log \textsl{g} $\le$ 5.5, and -1.0 $\le$ [M/H] $\le$ +0.5 in steps of 100 K, 0.5 dex, and 0.5 dex respectively. 

We interpolated in this grid to obtain model atmospheres with arbitrary parameters for the analyses of the primaries.  All the models were calculated on the same optical depth scale at 12000 \AA. This feature, and the fineness of the grid in the effective temperature domain, allowed us to use a three dimensional polynomial interpolation over the logarithm of the atmospheric parameters for each depth. 

\subsection{Procedure}
We determined the parameters iron abundance, [Fe/H], microturbulence, $\xi$, and macroturbulence, $\eta$, directly for the primaries by fitting synthetic spectra to profiles of the 30 selected Fe \textsc{i} lines in the observed spectra. Our procedure was similar to that used by \citet{carlos04} and was a two step process. 

In the first step, we used all the line profiles as a constraint to determine the global [Fe/H], $\xi$, and $\eta$. As for the determination of the astrophysical line data, we used an adaptation of the $\chi^{2}$ minimization algorithm of Marquardt \citep{mar63, press86} to find which parameters yielded synthetic spectra that best fit the observed spectra. For each iteration by the algorithm in [Fe/H], a model atmosphere with the appropriate parameters was interpolated from the grid described in the previous subsection. The model \textsl{T}$_{eff}$ was determined from the (\bv) -- \textsl{T}$_{eff}$ relationship of \citet{ramirez05}. We chose (\bv), as opposed to (\textsl{V -- K}) for example, because of the availability of a homogeneous set of precise photometry in these bands for our target stars. The uncertainty in our adopted \textsl{T}$_{eff}$ values was estimated by propagating the errors in the photometry through the \citet{ramirez05} formula and adding this in quadrature with the uncertainty in the formula itself. The model log \textsl{g} value and uncertainty was determined by using \citet{bertelli94} isochrones as described in \citet{carlos04}. The model atmosphere [M/H] was set equal to [Fe/H]. The \textsl{V} magnitudes, \bv colors, and parallaxes needed to determine these parameters were taken from the Hipparcos catalog \citep{hipp, perryman97}. The color relationship and isochrones depend weakly on metallicity, so the \textsl{T}$_{eff}$ and log \textsl{g} were determined for each iteration in [Fe/H]. 

We generated synthetic spectra for each line profile region with an updated version (described in detail in \S 5.1) of the plane-parallel, LTE, stellar analysis computer code MOOG \citep{sneden73} and the given model atmosphere. We use the macroturbulence parameter to account for large-scale turbulent and rotational broadening. The synthetic spectra are convolved with an isotropic Gaussian profile with a full-width half-max, FHWM, equal to the combination of the macroturbulence parameter value and the measured instrument resolution (2.55 km s$^{-1}$). The synthetic spectra were then resampled to the pixel scale of the observed spectrum and compared to it. The spectral regions used in evaluating the fit were determined by the mask described above. Additionally, we attempted to minimize errors introduced by departures from local thermodynamic equilibrium (LTE) by ignoring points in the line cores that are more than 0.5 residual intensity units below the continuum \citep{carlos04}. Therefore, the exact regions that were fit vary slightly among the objects. The match between the synthetic spectra and observed spectra was evaluated by the fitting algorithm, new parameters selected if necessary, and the algorithm continued to iterate. This process continued until the parameters which minimized $\chi^{2}$ were found. 

Once the best fit to all the line profiles together was found, the model atmosphere parameters, $\xi$, and $\eta$ were fixed to their final values and the process was repeated to determine $\epsilon$(Fe) for each line. The abundance found for each line was compared with the solar iron abundance, 7.45, so that each line gave a differential abundance, [Fe/H]. The final [Fe/H] value is the mean of the distribution of the line abundances. 

Our calculated uncertainties for the derived [Fe/H] values consist of three components. The first component ($\sigma_{l}$) is the uncertainty in the mean of the line abundance distribution, the standard deviation divided by the square root of the number of lines used minus one. The second ($\sigma_{t}$) and third ($\sigma_{g}$) components were derived from the uncertainties in the stellar \textsl{T}$_{eff}$ and log \textsl{g} values respectively. We re-ran the analysis for each star with the \textsl{T}$_{eff}$ and log \textsl{g} parameters set to one sigma above and below the adopted values. This yielded a pair of derived [Fe/H] values for each parameter. The RMS deviation from the adopted [Fe/H] value for the two pairs were the second and third components. The three components were added in quadrature to yield the final uncertainty in our derived [Fe/H] values. 

Our derived parameters for the solar-similar primaries are give in Table ~\ref{tab:table3}. The internal uncertainties in the derived \textsl{T}$_{eff}$, log \textsl{g}, and [Fe/H], including the [Fe/H] error components, for each star are also given. The median uncertainty in [Fe/H] for the solar-similar primaries is 0.06 dex. The uncertainties in the microturbulence and macroturbulence were assumed to be 0.15 and 0.20 km s$^{-1}$ respectively for all the objects. 

\subsection{Comparison of Results}
The most common abundance analysis techniques can typically approach internal precisions of 10\% for the abundances of elements with easily observed spectral lines. However, independent analyses of the same star can give abundances that differ by amounts many times quoted uncertainties \citep{cayrel01}. This is of particular concern for our investigation because we are interested in determining not only the internal consistency of our abundance analyses of two very different samples of stars, solar-similar stars and M dwarfs, but also our external consistency with the results of other groups. To accomplish this, we determined iron abundances for 30 stars (referred to hereafter as the ``test sample'') that were analyzed by \citet{carlos04} and \citet{vf05}. We selected objects in common with both of these studies and that had a range of stellar parameters bracketing those of our sample. We made use of the spectra, which are of similar resolution and S/N as our spectra, available via the Spectroscopic Survey of Stars in the Solar Neighborhood (S$^4$N) website\footnote{http://hebe.as.utexas.edu/s4n/index.html} and analyzed them in an identical manner as our sample. Our derived [Fe/H] values for this test sample have a median internal uncertainty of 0.07 dex.

\begin{figure}
\epsscale{1.0}
\plotone{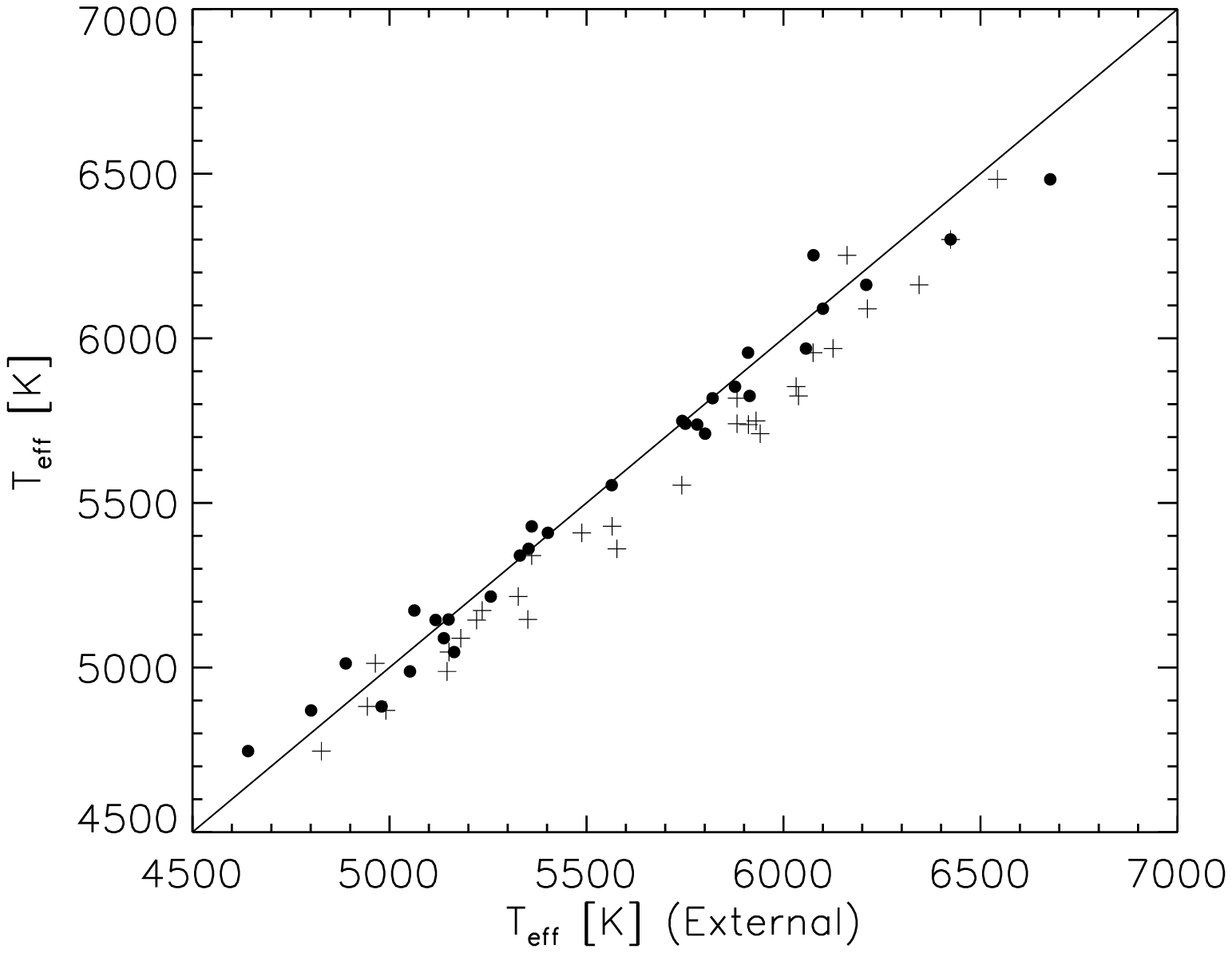}
\caption{Comparison of our determined \textsl{T}$_{eff}$ for the test sample with \citet{carlos04} (filled circles) and \citet{vf05} (plusses).}
\label{fig:f2}
\end{figure}

\begin{figure}
\epsscale{1.0}
\plotone{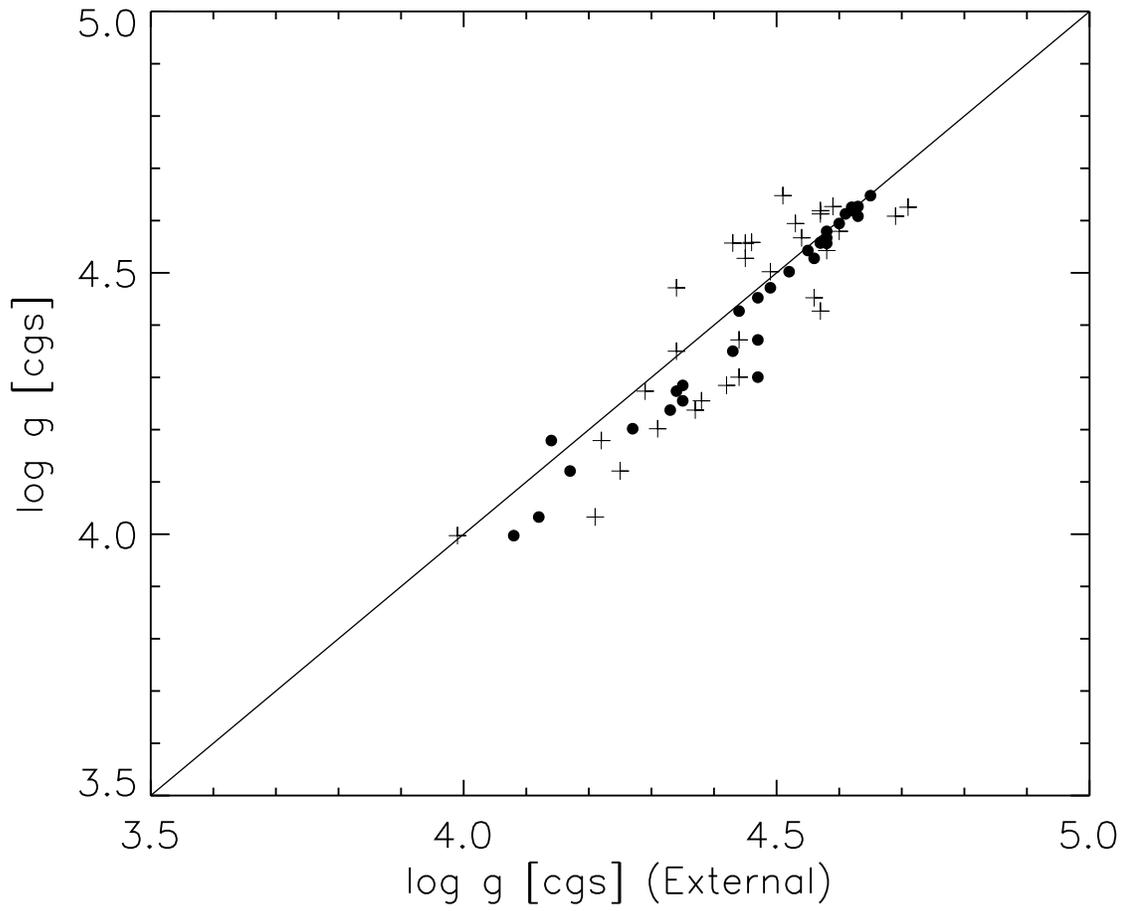}
\caption{Same as Figure ~\ref{fig:f2}, except for log \textsl{g}.}
\label{fig:f3}
\end{figure}

\begin{figure}
\epsscale{1.0}
\plotone{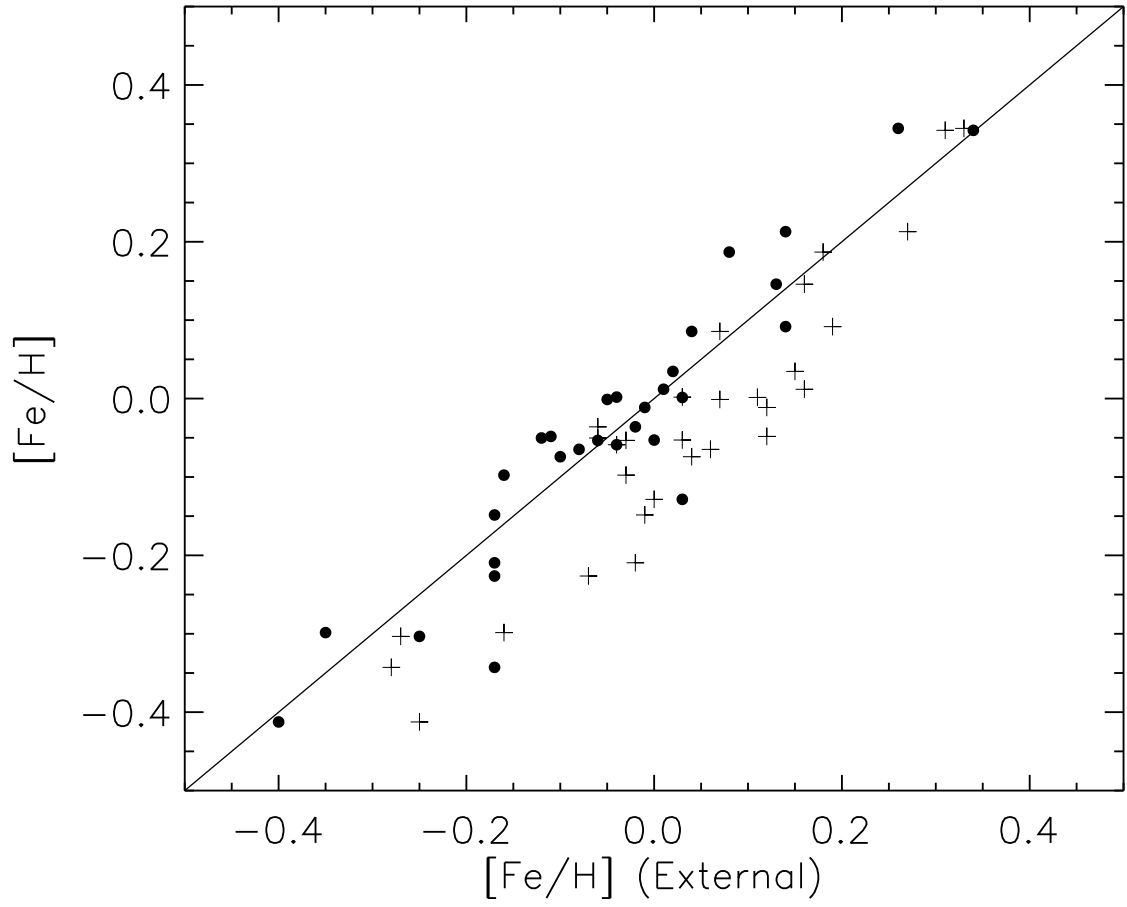}
\caption{Same as Figure ~\ref{fig:f2}, except for [Fe/H].}
\label{fig:f4}
\end{figure}

The derived \textsl{T}$_{eff}$, log \textsl{g}, and [Fe/H] values for the test sample from \citet{carlos04}, \citet{vf05}, and this study are collected in Table ~\ref{tab:table4}. Figures ~\ref{fig:f2}, ~\ref{fig:f3}, \& ~\ref{fig:f4} compare our derived \textsl{T}$_{eff}$, log \textsl{g}, and [Fe/H] for the test sample with those of \citet{carlos04} and \citet{vf05}. Both solve for model atmosphere [M/H] and [Fe/H] separately whereas we equate the two. Therefore we compare our [M/H] $\equiv$ [Fe/H] with their [Fe/H] values in Figure \ref{fig:f4}. 

For the 30 test sample stars, we find mean offsets ($X - X_{external}$, where ``$X$'' is a derived parameter) with respect to the results of \citet{carlos04} of -12 $\pm$ 15 K ($\sigma$ = 81 K), -0.04 $\pm$ 0.01 dex ($\sigma$ = 0.04 dex), and 0.00 $\pm$ 0.01 dex ($\sigma$ = 0.06 dex) for \textsl{T}$_{eff}$, log \textsl{g}, and [Fe/H] respectively. Good agreement for all the parameters is found due to the very similar analysis techniques. Compared to the results of \citet{vf05}, we find mean offsets of -117 $\pm$ 14 K ($\sigma$ = 75 K), -0.02 $\pm$ 0.02 dex ($\sigma$ = 0.10 dex), and -0.08 $\pm$ 0.01 dex ($\sigma$ = 0.07 dex) for \textsl{T}$_{eff}$, log \textsl{g}, and [Fe/H] respectively. Our \textsl{T}$_{eff}$ and [Fe/H] values for the test sample are systematically lower than their derived values over the entire range of the sample. Generally, this may be explained by systematic differences in techniques. The \citet{vf05} analysis is purely spectroscopic, while ours relies on outside constraints of the stellar \textsl{T}$_{eff}$ and log \textsl{g}. However, the exact reason for the systematic discrepancies is unknown. While detailed study of the causes of inter-study systematic differences is a worthwhile pursuit, it is beyond the scope of this investigation. 

Overall, the results of our analysis of the test sample are broadly consistent with the results of \citet{carlos04} and \citet{vf05} despite the differences in techniques and model atmospheres used. We conclude that our results for the primaries in the binary sample are robust against major external errors and precise enough to check for internal consistency with the results of our analysis of the M dwarf secondaries.

\section{APPLICATION OF THE V98 TECHNIQUE}
Our initial goal was to determine whether the V98 approach used with existing model atmospheres yielded metallicities for M dwarfs that were consistent with those derived for solar-similar stars with well established methods. We did this by analyzing the secondaries in our binary sample and comparing the derived metallicities with those that we determined for the corresponding primaries. In this section we discuss our application of this technique and the necessary details. We refer the reader to the V98 paper for more specifics.

In the V98 approach, synthetic spectra are matched to two observed spectral regions of an M dwarf spectrum. One region contains strong atomic lines (8670 -- 8700 \AA) and another contains a TiO bandhead (7078 -- 7103 \AA). To duplicate this method as closely as possible, we synthesized a grid of spectra for the two spectral regions using a version of SYNTH \citep{piskunov92} that was modified to handle the molecular equation of state as described by V98, NextGen model atmospheres \citep{haus99}, and the V98 line lists. Our grids had an increased \textsl{T}$_{eff}$ range, 3000 -- 4000 K in steps of 100 K, compared to V98 in order to cover the anticipated range of our sample. Microturbulence was set to zero during the synthesis of the grid. Our implementation used the same $\chi^{2}$ minimization algorithm that was used in our analysis of the primaries to determine the astrophysical parameters, \textsl{T}$_{eff}$, log \textsl{g}, [M/H], and $\eta$ which gave the best match between the synthetic and observed spectra. Additionally, as in V98, we included two continuum normalization factors, a zero-point and slope, per analyzed echelle order as free parameters in the fit. There were eight fit parameters in total.

The spectral points used to constrain the fit include the entire TiO bandhead region and the atomic line profiles to begin with. For a particular iteration in the parameters of the minimization function, interpolation in the synthetic spectrum grid using the same procedure as V98 yielded synthetic spectra for the values of \textsl{T}$_{eff}$, log \textsl{g}, and [M/H]. Macroturbulent and instrumental broadening was added after interpolation by convolution with a Gaussian. After the first best-fit was found, points in the bandhead region where the fit deviated from the observed spectrum by 0.15 residual intensity units were flagged to be ignored and the fit was re-run. The process continued iteratively as the rejection limit was decreased to 0.04 residual intensity units in steps of 0.01. The final parameters were arrived at after the last rejection iteration.

We analyzed the spectra taken for the M dwarf secondaries in the binary sample using our adaptation of the V98 technique. We found that the metallicities derived in this manner average 0.56 dex lower than the [Fe/H] values of the primaries. We considered that this systematic inconsistency could be reduced by improvements in the analysis method and model atmospheres as described in the next section.

\section{MODIFICATIONS TO THE V98 APPROACH}
Motivated by the inconsistency found above, we made a systematic study of the V98 technique. We implemented some potential improvements in our spectrum synthesis code, cool-star model atmospheres, and analysis technique. We also derived empirical surface gravity and abundance trend relationships which we used to constrain our analysis. The following subsections detail our work in this area.

\subsection{Spectrum Synthesis Code}
Following V98, we used SYNTH2 to generate synthetic spectra in our original application described in the previous section. For the new analysis we chose to use MOOG \citep{sneden73} for spectrum synthesis. We have modified MOOG to extend its capabilities to the M dwarf domain, while also maintaining consistency with its pre-existing functionality. This allowed us to use MOOG for the analysis of the solar-similar primaries and M dwarf secondaries.

We altered MOOG so that the chemical equilibrium calculations were carried out for an extensive set of molecules and atoms. We identified the set of species to include by examining the PHOENIX version 13.13.00E partial pressure tables. We selected all molecular species that have partial pressures greater than $10^{-7}$\% of the total gas pressure above 1500 K. The atomic species included were the neutral and singly ionized species of any elements that are a part of the selected molecules or needed specifically for continuous opacities. The equilibrium calculation included 16 different elements and 40 molecules for a total of 72 species. 

We also updated the molecular data in MOOG. Using the PHOENIX partial pressure tables, we fit fourth order polynomials as a function of $\theta \equiv$ log (5040/T) to each molecular species' partial pressure to construct equilibrium constants. We adopted the dissociation energies from \citet{sauval84}. The equilibrium constant fits are valid over the range in 1500 $\le$ T $\le$ 10000 K. We checked the equilibrium constants against the direct calculations from the partition functions of \citet{sauval84}. No major discrepancies were found and we adopted the equilibrium constants from the fits to the PHOENIX partial pressure data to maintain consistency with the model atmospheres we used in our analysis. 

Molecular energy level populations, which are needed to calculate line opacities, are normally calculated from the molecule's constituent elements' partition functions in MOOG. We chose to use the molecular partition function directly for TiO, the only molecular species included in our line list for spectrum synthesis, to maintain consistency with the molecular equilibrium calculations. We modified MOOG to use the partition function for $^{48}$Ti$^{16}$O (the dominant isotope) calculated by \citet{kurucz99}, which agrees well with that given by \citet{sauval84}.

We also adopted the TiO dissociation energy, $D_{0}$ = 6.87 $eV$, from the lab measurement of \citet{naulin97}. This is the same value as given by \citet{huber79}, \citet{sauval84}, and the JANAF Thermochemical Tables \citep{janaf85}, but 0.06 $eV$ lower than the lab measurement of \citet{dubois77}.

\subsection{Model Atmospheres}
We computed a new grid of model atmospheres with PHOENIX (version 13) for our revised M dwarf analysis. These models are an  updated version relative to the NextGen version models used by V98 and in \S 4. Important updates and revisions since the NextGen models were released are discussed by \citet{kucinskas05}. Additionally, we used the TiO partition function mentioned in the previous subsection, which is roughly a factor of three lower than that used in all previous versions of PHOENIX. We also used the recently updated solar abundances presented in the compilation by \citet{asplund05}. Most relevant for this work was the reduction of the solar abundances of C, O, and Ti, by 0.17, 0.21, and 0.12 dex respectively from the values used in previous versions. Also, all of the model atmospheres were computed assuming a microturbulent velocity of 2 km s$^{-1}$. The new model grid spans the range in parameters 3000 $\le$ \textsl{T}$_{eff}$ $\le$ 4000 K, 4.5 $\le$ log \textsl{g} $\le$ 5.5, and -1.0 $\le$ [M/H] $\le$ +0.5 in steps of 100 K, 0.5 dex, and 0.5 dex respectively.

\subsection{TiO Line Data}
One key aspect of the V98 analysis was their inclusion of a strong TiO bandhead, $\gamma$ R$_{2}$ 0 -- 0, as a fit constraint. While molecular bandheads typically have good temperature sensitivity, they are complex and require significant consideration of line data for proper synthesis. Additionally, a minimum level of TiO line ``haze'' exists throughout the visible and near-IR regions of M dwarf spectra. This necessitates inclusion of TiO lines for realistic spectrum synthesis in the atomic line spectral regions that we include in our analysis. We adopted the TiO line list of \citet{plez98}, which is based on the most recent laboratory measurements and theoretical calculations. We chose to use the version of the list that has the laboratory line positions substituted for the calculated values where they are available. We used the formula given by \citet{schweitzer96} to calculate the van der Waals damping for molecular lines.

The TiO line list of \citet{plez98} includes data for 15.7 million lines and includes many more lines than have a meaningful impact on the spectra of M dwarfs. To save computational time, we used a similar procedure as V98 to determine a strength cutoff for lines to be included in our line lists. We calculated the strength, S, based on the formula given in V98, for all the lines in the three spectral regions we use in our analysis. The formula is: \begin{equation} S = log(agf\lambda) - \theta\chi,\end{equation} where $a$ is the isotopic abundance fraction, $g$ the statistical weight, $f$ the oscillator strength, $\lambda$ the wavelength, and $\chi$ the lower level excitation energy.

For each of the three regions separately, we progressively lowered the strength cutoff value from the maximum value by 0.5, until spectra generated for a characteristic model atmosphere did not change by more than 0.5\%. This resulted in strength cutoff values of -0.30, -0.51, and -0.26 for the spectral regions 7080 -- 7100, 8320 -- 8430, and 8650 -- 8700 \AA\ 
and yielded 16,667, 35,695, and 14,487 TiO lines respectively. A comparison of spectra synthesized using MOOG with the V98 line list and the new line list showed that the revised log \textsl{gf}s in new list yielded smaller depths for the 7088 \AA\ bandhead at all model temperatures. For unsmoothed spectra generated with a model atmosphere having \textsl{T}$_{eff}$ = 3500 K, the V98 line list gave residual intensities an average of 0.07 lower than the new line list.

\subsection{Atomic Line Data}
In addition to the TiO bandhead around 7088 \AA, V98 focused on fitting five relatively strong atomic lines in a second spectral interval, 8670 -- 8700 \AA. As part of our modification to the original technique, we have expanded this list of lines to include 11 more. 

We examined our observed spectra for new lines that met our search criteria, which was similar to that used in V98. Our first criterion was that the new lines must have fractionally small amounts of TiO line blending. This necessitated the use of fairly strong lines as a minimum level of molecular line haze exists throughout the observed spectral regions. The second criterion was that they must also be strong enough in a solar spectrum for the purpose of determining astrophysical log \textsl{gf}s. 

The new lines we selected occupy a spectral interval adjacent to the original one and in the same echelle order, 8650 -- 8670 \AA, and a second echelle order, 8320 - 8430 \AA, in our observed data. This makes for convenient spectral synthesis and increases the constraint on the continuum normalization, which introduces two free parameters into the fit for each echelle order utilized. The new lines also include a new elemental species, Ca. One of the lines added is a member of the Ca \textsc{ii} ``infrared triplet.'' 

For consistency, we determined astrophysical log \textsl{gf}s for the lines originally used by V98 and the new ones used here. We did this using the same procedure described in \S 3.1. Eleven of the lines have accurate van der Waals damping data from \citet{barklem00}, which we took advantage of. For the remaining five lines we used the approximation of \citet{unsold55} enhanced by a factor of 2.5 as was done by V98. The final data for the 16 lines is given in Table ~\ref{tab:table5}. Our log \textsl{gf} values are essentially identical with the V98 values for the two Fe \textsc{i} lines originally used. The log \textsl{gf}s for the Ti lines are different from the V98 values by larger amounts due to the 0.12 dex lower value for the solar Ti abundance that we adopted. Also, two of the Fe \textsc{i} lines that we use to analyze the M dwarfs were also used in the analysis of the solar-similar primaries.

We constructed an atomic line list for our M dwarfs using data obtained from the Vienna Atomic Line Database \citep[VALD, ][]{piskunov95, kupka99}. We queried VALD for lines in the three spectral regions we analyze with assumed model temperatures of 3000 and 4000 K. The lists for each temperature were merged and we substituted our astrophysical \textsl{gf}s for the 16 lines we fit. These atomic line lists were then combined with the TiO line lists mentioned above to create the final line lists for our analysis.

\subsection{Microturbulence}
The use of a variable parameter representing a depth independent, small scale (relative to a photon mean free path) velocity distribution, microturbulence, is standard procedure in high precision, one-dimensional spectroscopic analyses. Following convention, we allowed microturbulence to vary in our analysis of the primaries in the binary sample. Also, we elected to introduce a variable microturbulence in our analysis of the M dwarfs. Microturbulence, unlike macroturbulence, must be accounted for in the spectral synthesis itself and cannot be be added by a later convolution. Therefore, a pre-synthesized grid of spectra for our purpose would have to include a range of values for a fourth parameter. We chose to abandon this approach as we had little indication of what the microturbulence values would actually be, and because the already significant grid creation time would be multiplied by the number of microturbulence grid nodes. 

Our revised technique was similar to that used in the analysis of the solar-similar primaries. A new model atmosphere was interpolated from the grid mentioned in \S 5.2 using the interpolation method described in \S 3.2 for each iteration in the stellar parameters by the $\chi^{2}$ minimization algorithm. With that model atmosphere and the microturbulence suggested by the fitting program as inputs, MOOG was used to generate a synthetic spectrum. Macroturbulence and instrumental broadening were added by a convolution at this point. Then the synthetic spectra were resampled to the wavelength scale of the spectrograph and compared to the observed spectra. We note that the individual microturbulences determined for the analyzed stars are lower by $\sim$ 1 km s$^{-1}$ from the value used to generate the model atmospheres. However, a change of the microturbulence adopted for generating the model atmospheres of this magnitude would not have a noticeable impact on their structure.

\subsection{Surface Gravity}
All spectroscopic analyses can benefit from outside constraint on any of the needed stellar parameters. In the case of M dwarfs, current theoretical models are not yet to the point where they can be used for this purpose. Fortunately, increasing attention has been focused on measuring the physical properties of M dwarfs since the V98 analysis. Specifically, the direct measurements of M dwarf masses, $\mathcal{M}$, and radii, $\mathcal{R}$, have permitted the calculation of empirical surface gravities for a sample of M dwarfs. 

We have deviated from the purely spectroscopic approach of V98 by calculating and adopting an empirical log \textsl{g} -- $\mathcal{M}$ relationship. We compiled all the known M dwarf radii measurements with precisions better than 16\%. The measurements come from the observations of M dwarfs in eclipsing binary systems \citep{metcalfe96, ribas03, torres02, maceroni04, maxted04, creevey05, lopez05} and interferometric observations of single M dwarfs \citep{lane01, seg03, berger06}. We also compiled the high precision mass measurements for the stars in the eclipsing binary systems. We calculated the masses for the single stars from the M$_{K}$ -- $\mathcal{M}$ relationship in \citet{delfosse00}. The \textsl{K} magnitudes and parallaxes needed to calculated the M$_{K}$ values were taken from the Two Micron All Sky Survey (2MASS) point source catalog \citep{cutri03} and the Hipparcos catalog respectively. Uncertainties in the masses calculated in this manner were assumed to be 10\%. We then calculated log \textsl{g} for each object from the mass and radius data. Uncertainties in both mass and radius were propagated through formula to give the the associated uncertainty in log \textsl{g}. 

\begin{figure}
\epsscale{1.0}
\plotone{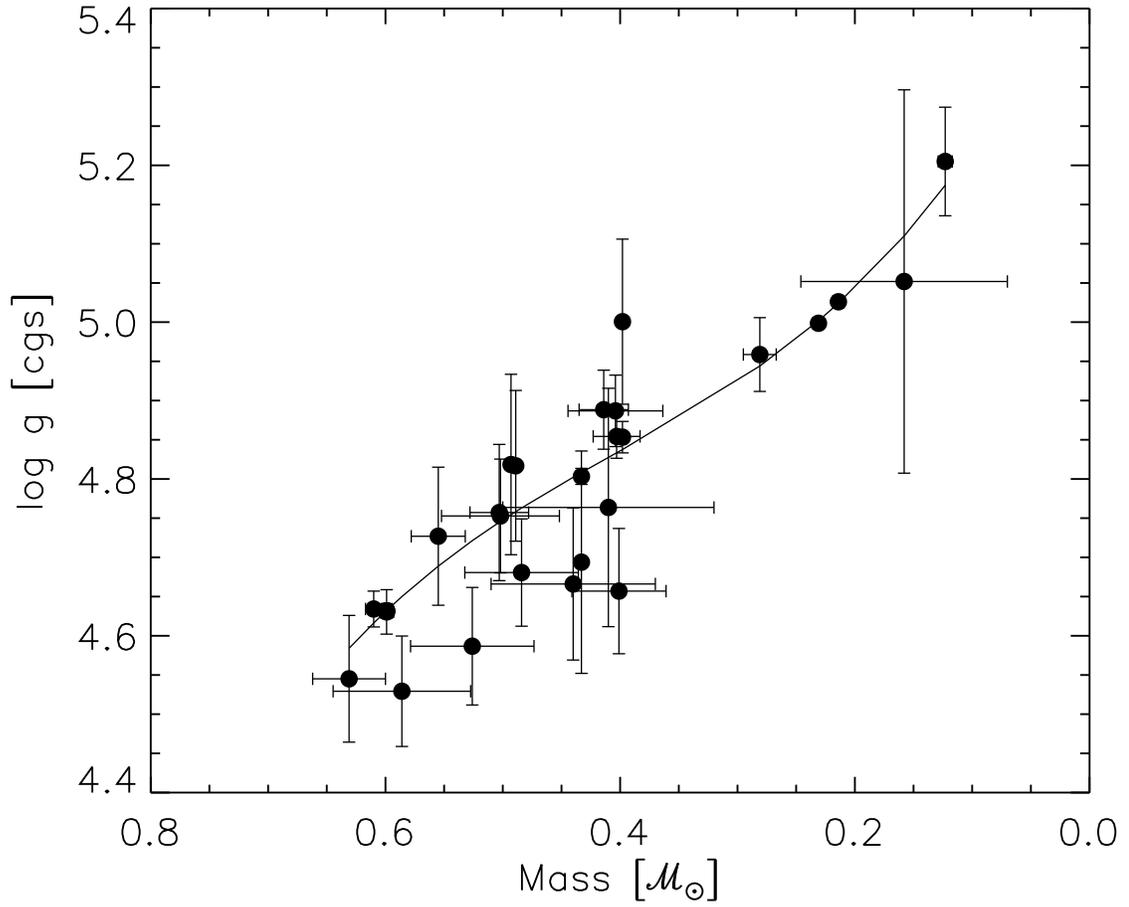}
\caption{Plot of the empirical log \textsl{g} and $\mathcal{M}$ data (points) and the fit given in equation (2) (solid line).}
\label{fig:f5}
\end{figure}

With this this dataset, we fit a third order polynomial to the surface gravity values as a function of mass. The data points were weighted according to their uncertainties in both parameters. Twenty-eight independent points were considered in the fit, with values spanning the range in mass: 0.123 -- 0.621 $\mathcal{M}_{\sun}$ and radius: 0.145 -- 0.702 $\mathcal{R}_{\sun}$. The function found was \begin{equation} log \: g = 5.491 - 3.229 \mathcal{M}_{\star} + 5.949 \mathcal{M}_{\star}^{2} - 4.929 \mathcal{M}_{\star}^{3}, \end{equation} where log \textsl{g} is in cgs units and $\mathcal{M}$ is in solar units. The standard deviation for this relationship is 0.08 dex and the data and fit are plotted in Figure ~\ref{fig:f5}.

Our revised technique for analyzing M dwarfs makes use of equation (2) to fix log \textsl{g}. For the M dwarfs in our binary sample, we calculated M$_{K}$ magnitudes based on 2MASS photometry and Hipparcos parallaxes. We converted those values to masses based on the M$_{K}$ -- $\mathcal{M}$ relationship in \citet{delfosse00} and then used these masses to estimate log \textsl{g} from equation (2). The theoretical log \textsl{g} -- $\mathcal{M}$ and the M$_{K}$ -- $\mathcal{M}$ relationships are independent of stellar metallicity, unlike the M$_{V}$ -- $\mathcal{M}$ relationship \citep{baraffe98}. Therefore the use of both relationships is appropriate for our purposes. This approach eliminates one parameter to be determined from the fitting process and permitted a more robust test of the model atmospheres.

\subsection{Abundance Trends}
V98 fit synthetic spectra to various spectral features to derive a single abundance metric, metallicity. The shapes of these features are directly related to the abundances of three species, O, Ti, and Fe, and indirectly to the abundances of many other species through their affect on the chemical equilibrium. We retained this approach and added a Ca line as a further constraint. The determination of the individual abundances of these elements is not possible at this point as the varied features are needed to break the degeneracies between the parameters that we determined from the spectra ({T}$_{eff}$, [M/H], $\xi$, $\eta$, and 6 continuum normalization factors). Given this, we had to adopt relationships between the abundances of various elements, [X/H], and our varying abundance parameter, [M/H]. 

First, we assumed [Fe/H] = [M/H], as we also did for the solar-similar primaries. Then we calculated relationships between alpha element and carbon abundances and [Fe/H] using the abundance data of \citet{carlos04}. For the alpha elements, we combined the elemental abundances of O, Mg, Si, Ca, and Ti into a single data set. Ti is not strictly an alpha element, but the observed trend of its abundance with [Fe/H] closely matched the other elements in the range of values covered so we included it. We fit these elemental abundances with a single parameter, [$\alpha$/H], as a linear function of [Fe/H]. The expression found was \begin{equation} [\alpha/\mathrm{H}] = 0.11 + 0.87 [\mathrm{Fe/H}]\end{equation} for 607 data points with a standard deviation of 0.10 dex. This formula obviously gives super-solar abundances for the alpha elements at [Fe/H] = 0 and, as \citet{carlos04} noted, suggests that either the Sun has an unusual abundance pattern compared to solar-neighborhood stars or that the data is affected by systematic errors. Adopting this formula means that our analysis was not strictly differential to the Sun. Rather, our revised method was differential to an artificial [M/H] = 0 model with mean solar neighborhood abundances. This approach was not ideal, but was necessary to maximize our ability to analyze a sample of stars, rather than a single individual star.

\begin{figure}
\epsscale{1.0}
\plotone{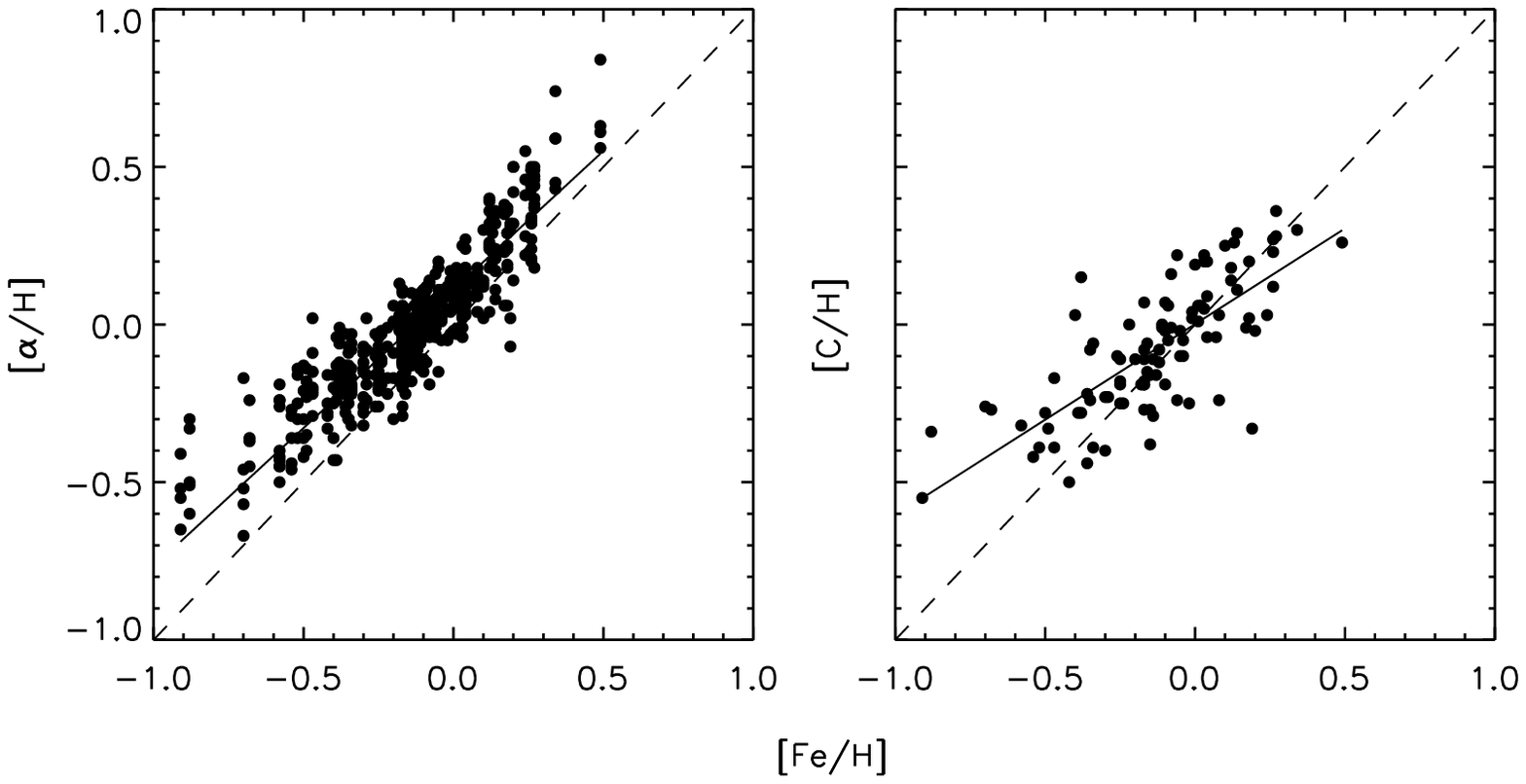}
\caption{Alpha element (left panel) and carbon (right panel) abundances versus [Fe/H] (points) from \citet{carlos04}. Error bars are omitted for clarity. The median uncertainties for the alpha element and iron abundances are 0.05 dex and 0.06 dex respectively. Carbon abundance uncertainties were set to 0.20 dex. The fits (solid lines) given in equations (3) and (4) and [X/H] = [Fe/H] relationships (dashed lines) are shown for comparison.}
\label{fig:f6}
\end{figure}

We also constructed a dataset of C abundances and fit [C/H] as a linear function of [Fe/H]. The resulting formula for 101 data points is \begin{equation} [\mathrm{C/H}] = -0.01 + 0.62 [\mathrm{Fe/H}]\end{equation} and has a standard deviation of 0.14 dex. The alpha element and carbon abundance data are plotted as functions of [Fe/H] in Figure ~\ref{fig:f6} along with the fits to the data given in equations (3) and (4).

For all other elements we assumed [X/H] = [Fe/H]. Of these, Na is particularly important because it is the dominant electron donor in M dwarf atmospheres. It was found that [Na/Fe] $\approx$ 0 for [Fe/H] $>$ -1.0 by \citet{reddy03}.

\section{RESULTS AND ERROR ANALYSIS}
\begin{figure}
\epsscale{1.0}
\plotone{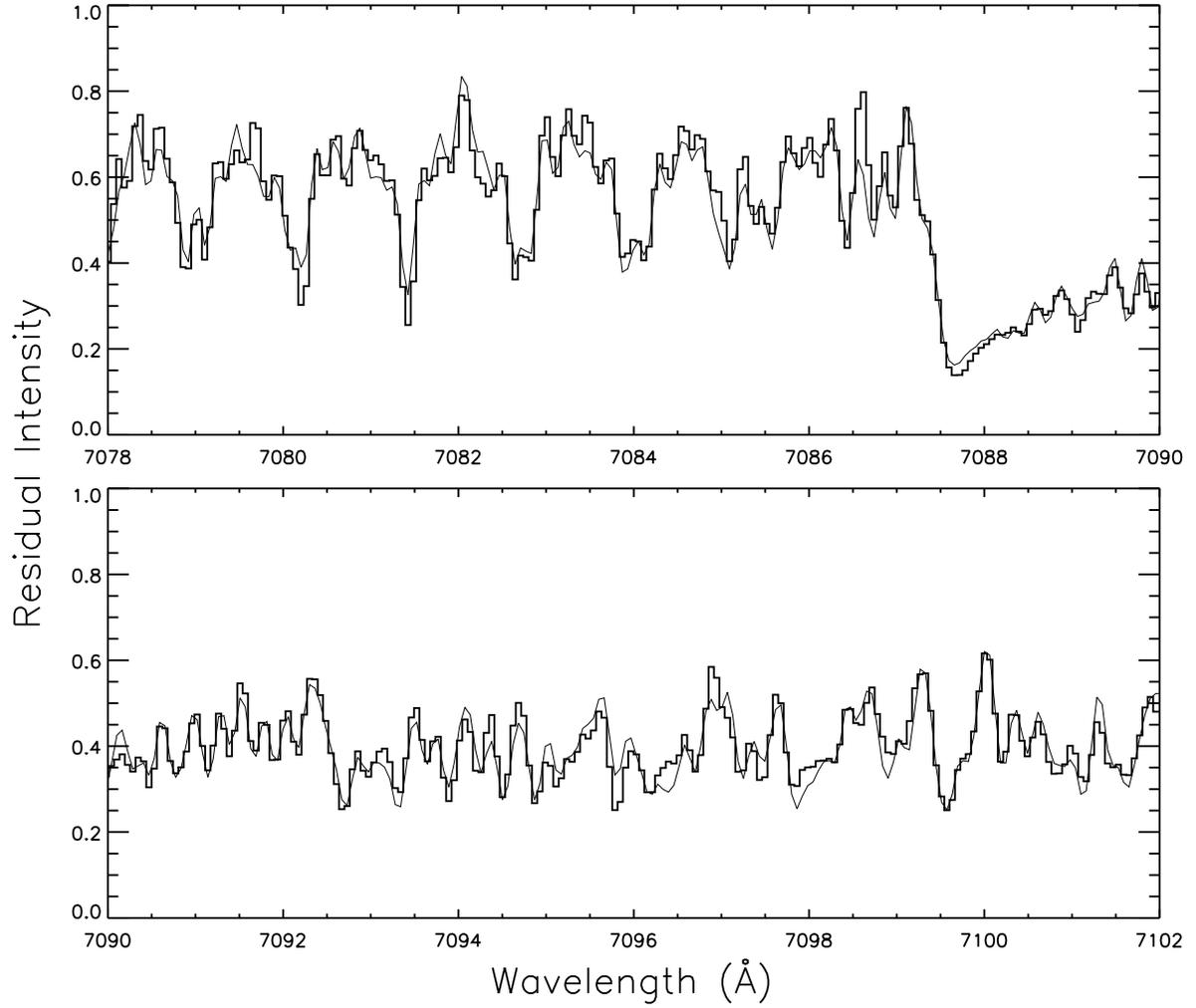}
\caption{Spectral region near the strong TiO $\gamma$ R$_{2}$ 0 -- 0 bandhead for HIP 12114B (histogram). The best fit used to determine the stellar parameters is over-plotted (solid line).}
\label{fig:f7}
\end{figure}

\begin{figure}
\epsscale{1.0}
\plotone{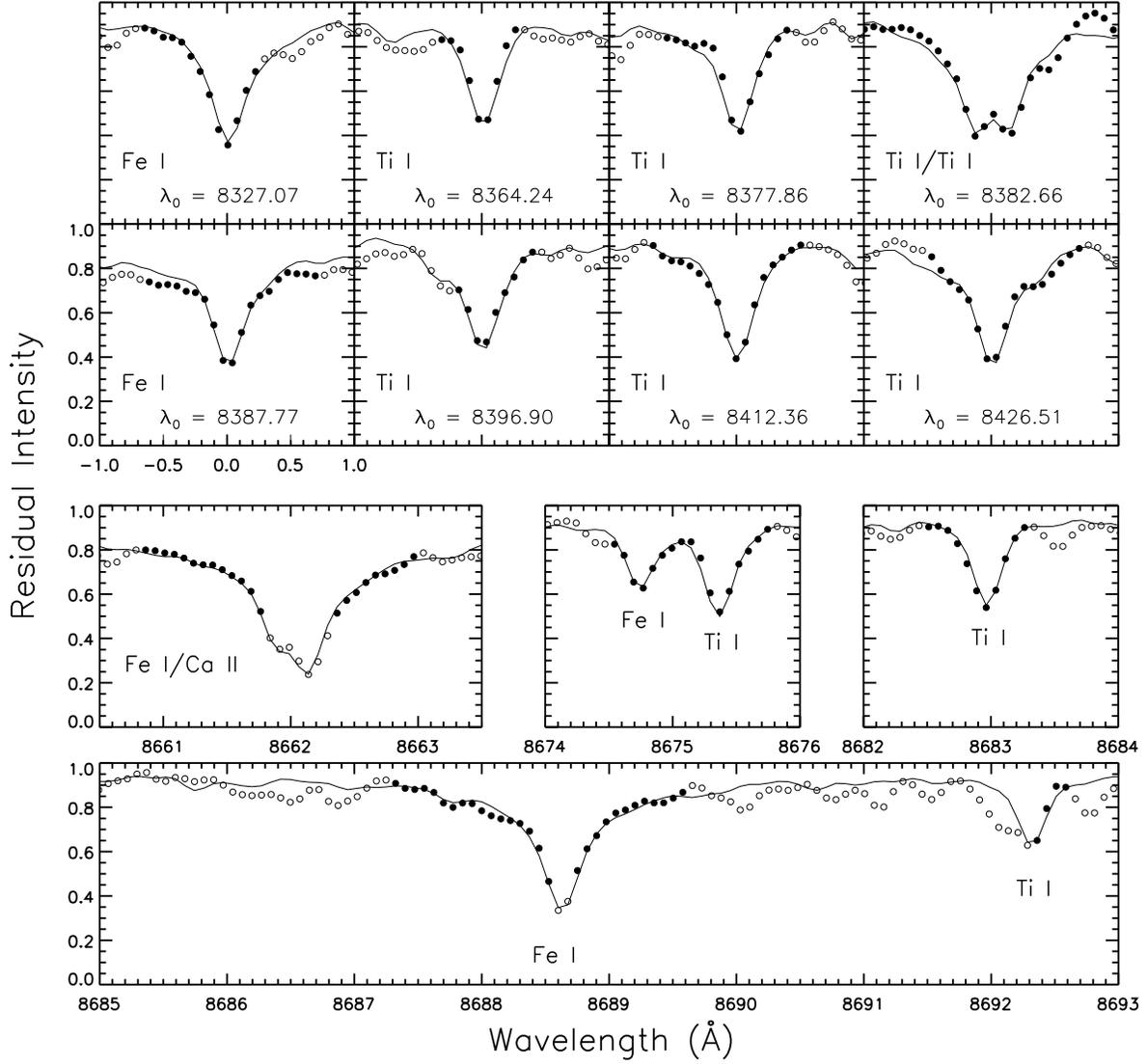}
\caption{Fit of synthetic spectra (solid line) to atomic line profiles (points) for HIP 12114B. The filled points were used in the fitting process; the open points were ignored. The panels are sorted by wavelength and the linear scaling in both parameters is the same throughout. The lines in each half, top and bottom, make up a contiguous spectral order in our observed spectra. The lines from 8674 -- 8693 \AA\ were used originally by V98; the others were added in this study. All apparent ``lines'' in the figure that aren't fit are actually multiple TiO lines.}
\label{fig:f8}
\end{figure}

The results of our analysis of the M dwarf secondaries using the revised technique described in \S 5 are given in Table ~\ref{tab:table6}. Plots of the observed spectrum and best fit for the coolest M dwarf, HIP 12114B, are shown in Figures ~\ref{fig:f7} and ~\ref{fig:f8}. The mean offset of the derived metallicities for the five M dwarfs from the expected values as given by their corresponding solar-similar companion ([M/H]$_{secondary}$ $-$ [M/H]$_{primary}$) is -0.08 $\pm$ 0.04 dex ($\sigma$ = 0.07 dex). 

The binary sample can be used for evaluating the errors in our technique because the derived metallicities are independent of the values derived for the primaries. We calculated the RMS deviation for the sample metallicities to be 0.11 dex by adding the average offset and it's standard deviation in quadrature. Assuming the systematic offset is a product of random errors and a symmetrical distribution, the RMS value is the standard deviation of a Gaussian probability distribution with a mean of zero. We added this value in quadrature to the median uncertainty in our derived metallicities for the primaries metallicities, 0.06 dex, to get 0.12 dex. We adopt this as the uncertainty in [M/H] for our technique.

With respect to the log \textsl{g} values determined from our empirical log \textsl{g} -- $\mathcal{M}$ relationship, we considered the uncertainty in the empirical relationship and the estimated masses needed for it. We assumed errors of 10\% in the estimated masses from the empirical relationship of \citet{delfosse00}. Propagating that through equation (2) yielded associated uncertainties of 0.06 dex in log \textsl{g}. Adding that value in quadrature to the uncertainty of equation (2), 0.08 dex, yielded 0.10 dex, our estimated uncertainty in the log \textsl{g} values for our M dwarfs.

The primary constraint of the stellar {T}$_{eff}$ in our analysis technique comes from the TiO bandhead used in the fitting process. Molecular bandheads in general are very sensitive to the temperature and composition, but not the gas pressure of the physical environments in which they form. We therefore used the uncertainty in our derived [M/H] values, which were determined externally, to estimate the errors in our derived {T}$_{eff}$ values. We repeated our analysis of the five M dwarfs with [M/H] fixed to values 0.12 dex above and below the best fit value. The parameters {T}$_{eff}$, $\xi$, and $\eta$ were allowed to vary and log \textsl{g} was fixed to the value determined from equation (2). We calculated the deviations of the five pairs of {T}$_{eff}$ values from the best fit values give in Table ~\ref{tab:table6}. The average of the 10 deviations is 48 K and we adopt this as the standard uncertainty in our derived {T}$_{eff}$ values. As a check of the independence of the {T}$_{eff}$ and log \textsl{g} parameters in our analysis, we repeated the above procedure with log \textsl{g} fixed to values 0.10 dex above and below that given by equation (2). We found an average deviation for {T}$_{eff}$ of 4 K, an insignificant change which supports our supposition.

Finally, we adopted the same standard uncertainties in our derived microturbulence and macroturbulence values as for the primaries, 0.15 and 0.20 km s$^{-1}$ respectively. 

\section{DISCUSSION}
We have carried out a test of the V98 technique to determine M dwarf metallicities using an analysis of binary star pairs. Our result is that the V98 technique yielded values that were systematically too low. Motivated by this, we have made modifications to the original technique and a re-analysis of the M dwarfs secondaries validates our new approach. Our modifications include expanding the stellar analysis code MOOG, adopting relationships for the abundances of the alpha elements and carbon to iron, using TiO line data from \citet{plez98}, inclusion of new atomic lines and data, introducing a variable microturbulence, and the use of an empirical surface gravity relationship. We have also made improvements in the PHOENIX model atmospheres that could have implications beyond those noted in our focused study. 

Our modified technique, in conjunction with new model atmospheres, has yielded a technique for determining M dwarf metallicities consistent to 0.11 dex with the techniques applied to solar-similar stars. We have assumed that the mean offset in derived metallicities between the primary and M dwarf parent samples is actually zero and that the observed offset of -0.08 $\pm$ 0.04 dex is due to random errors. The observed offset is actually 2$\sigma$ from zero and might be indicative of an unknown systematic in our technique or the model atmospheres. We estimated the uncertainties in our derived stellar parameters for the M dwarfs to be 48 K, 0.10 dex, 0.12 dex, 0.15 km s$^{-1}$, and 0.20 km s$^{-1}$ for {T}$_{eff}$, log \textsl{g}, [M/H], $\xi$, and $\eta$ respectively based on the assumption of no systematic errors. In future studies of single M dwarfs, we will adopt these values as the standard uncertainties in our analysis method when used with the particular model atmospheres described here.

Although we find good results for our study of the five binary pairs presented here, we are aware of several issues that could affect the results of our analyses of the primaries and the M dwarfs. In our analysis of the solar-similar primaries, there are potential systematics introduced by using the (\bv) -- \textsl{T}$_{eff}$ relationship of \citet{ramirez05}. Applying their relationship to the Sun using the most probable (\bv)$_{\sun}$ of 0.64 \citep{holmberg06} yields a \textsl{T}$_{eff}$ of 5706 K, which is 71 K lower than the true \textsl{T}$_{eff\sun}$. We have estimated the errors in our derived metallicities for the primaries due to uncertainties in our adopted \textsl{T}$_{eff}$ values on the same order as this difference (\S 3.3), but our approach ignores the potential systematics in the \citet{ramirez05} relationship for which it could be evidence. Nonetheless, we consider the \textsl{T}$_{eff}$ values derived from the \citet{ramirez05} relationship to be preferable to spectroscopically determined values for solar type stars which have their own potential problems \citep{ramirez04}.

Additionally, the model atmospheres we used in our analysis of the the primaries used the ``old'' solar abundance values given by \citet{anders89} and were not consistent with the models we used to analyze the M dwarfs. However, the abundances of the species that differed the most between the model versions used, C and O, affect the models of the primaries much less than the M dwarfs models. Model atmospheres of M dwarfs are heavily influenced by molecular opacities which depend directly and indirectly on the abundances of C and O. In comparison, the 1D model atmosphere structures of solar-similar stars are only minimally affected by opacities related to the C and O abundances at the level at which they have been altered (0.17 and 0.21 dex respectively).

Our technique for deriving the stellar parameters for M dwarfs relies on a number of assumptions that are not needed in analyses of solar-similar stars. These assumption could potentially introduce systematic errors that are not obvious given our test sample size. Improvements in our technique, including determining the abundances of specific elements instead of just the global metallicity, would be possible if there were more robust ways to estimate the stellar {T}$_{eff}$ and log \textsl{g} values as there are for solar-type stars. 

Our derived {T}$_{eff}$ values for the M dwarfs follow the expected trend and decrease smoothly with spectral type. However, when comparing these values with the {T}$_{eff}$ values as functions of spectral type in Table 4.1 of \citet{reidhawley04}, there is a systematic trend of increasing deviation with spectral type. For the earliest M dwarf in our sample (HIP 32423B, M0.5), there is essentially no difference in our derived {T}$_{eff}$ and that given by interpolating in Table 4.1. For our sample, the deviation increases linearly with spectral type and is roughly 300 K for the latest M dwarf in our sample (HIP 12114B, M3.5). This comparison assumes that there is minimal dispersion in the spectral type -- {T}$_{eff}$ relationship, which is only true at a rough level (N. Reid, private communication) as such a relationship ignores the effects of metallicity. Also, only two M dwarfs, GJ 699 \citep{dawson04} and GJ 411 \citep{leggett96, seg03} have known strictly empirical {T}$_{eff}$ values based on measured bolometric fluxes and radii. However, the suggestion of a systematic error in our derived temperature scale should not be ignored lightly. As we have shown in \S 6, changes in our adopted stellar {T}$_{eff}$ values of $\sim$ 50 K correspond to differences of 0.12 dex in [M/H]. We plan to apply our analysis to the M dwarfs with known strictly empirical {T}$_{eff}$ values in order to test the accuracy of our spectroscopically derived {T}$_{eff}$ values and attempt to resolve this issue.

Also, we use an empirical relationship to constrain the surface gravity of our M dwarfs. Adopting values derived from fully validated evolutionary models might be more reliable because it would be possible to include the effects of varying stellar ages. We suggest that the biggest opportunity to advance detailed spectroscopic analyses of M dwarfs would by improving low-mass stellar evolutionary models. All stellar structure models depend on realistic model atmospheres as a boundary condition. Therefore, the new PHOENIX model atmospheres that we have validated in our analysis could be used to produce more reliable evolutionary models. 

Although there is much left to do in the area of abundance analyses for M dwarfs, the work presented here has been the next step in the detailed study of these neglected objects. We consider our new technique to be robust enough for application to some current outstanding problems. Our research in the near-future will be focused on determining metallicities of M dwarfs with candidate extrasolar planets and those with accurate dynamical masses using the technique developed here.

\acknowledgments
We thank David Yong and David Doss for their assistance in taking the observations. We are indebted to Jeff Valenti for access to some of the computer programs used in the original analysis and for helpful discussions. Bertrand Plez kindly provided his TiO line list, for which we are grateful. We also thank David Lambert, Carlos Allende Prieto, and Barbara McArthur for their insight regarding this project. This research has made use of the SIMBAD database, operated at CDS, Strasbourg, France and the Vienna Atomic Line Database (VALD). GFB and JLB acknowledge support from NASA GO-06036, GO-06047, GO-06764, GO-08292, GO-08729, GO-08774, GO-09234, GO-09408, and GO-10773 from the Space Telescope Science Institute, which is operated by the Association of Universities for Research in Astronomy, Inc., under NASA contract NAS5-26555; and from JPL 1227563 ({\it SIM} MASSIF Key Project, Todd Henry, P.I.), administered by the Jet Propulsion Laboratory.

\clearpage
\begin{deluxetable}{ccccccc}
\tabletypesize{\scriptsize}
\tablecolumns{7}
\tablewidth{0pc}
\tablecaption{Observed binary pairs.}
\tablehead{
 \colhead{} & 
 \multicolumn{2}{c}{Primary (A)} &
 \multicolumn{2}{c}{Secondary (B)} \\
 \colhead{Name} & 
 \colhead{Spectral Type} & 
 \colhead{V} &
 \colhead{Spectral Type} & 
 \colhead{V} &
 \colhead{$\pi$ (mas)\tablenotemark{a}} &
 \colhead{$\rho$ (arcesec)\tablenotemark{b}}
}
\startdata
HIP 12114  & K3 V & 5.79 & M3.5 V   & 11.66 & 138.72 $\pm$ 1.04 & 165\tablenotemark{d} \\
HIP 26907  & K1 V & 8.56 & M  V\tablenotemark{c}   & 13.21 &  31.90 $\pm$ 1.28 & 53\tablenotemark{e} \\
HIP 32423  & K3 V & 8.80 & M0.5 V & 12.17 &  40.02 $\pm$ 1.22 & 31\tablenotemark{d} \\
HIP 40035  & F7 V & 5.53 & M2 V   & 12.26 &  44.47 $\pm$ 0.77 & 92\tablenotemark{d} \\
HIP 102040 & G5 V & 6.44 & M2.5 V & 11.80 &  47.65 $\pm$ 0.76 & 125\tablenotemark{d} \\
\enddata
\label{tab:table1}
\tablenotetext{a}{\citet{hipp, perryman97}.}
\tablenotetext{b}{Component separation on the sky.}
\tablenotetext{c}{Based on position in an HR diagram.}
\tablenotetext{d}{\citet{poveda94}.}
\tablenotetext{e}{\citet{allen00}.}
\end{deluxetable}

\clearpage
\begin{deluxetable}{cccc}
\tabletypesize{\scriptsize}
\tablecolumns{4}
\tablewidth{0pc}
\tablecaption{Central wavelengths ($\lambda$), lower level excitation potential ($\chi$), and lab and astrophysical log \textsl{gf}s for the Fe \textsc{i} lines used in the analysis of the primaries.}
\tablehead{
 \multicolumn{1}{c}{$\lambda$} &
 \multicolumn{1}{c}{$\chi$} &
 \multicolumn{2}{c}{log \textsl{gf}} \\
 \colhead{(\AA)} & 
 \colhead{(eV)} &
 \colhead{Lab} & 
 \colhead{Solar} \\ 
}
\startdata
 5956.6943 & 0.859 & -4.498 & -4.515 \\
 6065.4824 & 2.608 & -1.410 & -1.653 \\
 6079.0093 & 4.652 & -1.020 & -0.949 \\
 6082.7104 & 2.223 & -3.570 & -3.516 \\
 6085.2588 & 2.758 & -3.050 & -2.899 \\
 6093.6445 & 4.607 & -1.400 & -1.298 \\
 6096.6655 & 3.984 & -1.830 & -1.768 \\
 6127.9067 & 4.143 & -1.399 & -1.368 \\
 6229.2285 & 2.845 & -2.830 & -2.896 \\
 6232.6411 & 3.654 & -1.223 & -1.238 \\
 6240.6460 & 2.223 & -3.173 & -3.249 \\
 6246.3188 & 3.602 & -0.877 & -0.878 \\
 6252.5552 & 2.404 & -1.767 & -1.813 \\
 6265.1338 & 2.176 & -2.550 & -2.665 \\
 6270.2251 & 2.858 & -2.609 & -2.546 \\
 6322.6855 & 2.588 & -2.430 & -2.446 \\
 6411.6494 & 3.654 & -0.717 & -0.695 \\
 6430.8462 & 2.176 & -1.946 & -2.202 \\
 6481.8701 & 2.279 & -2.980 & -2.920 \\
 6498.9370 & 0.958 & -4.689 & -4.592 \\
 6593.8706 & 2.433 & -2.420 & -2.440 \\
 6597.5610 & 4.795 & -0.970 & -0.874 \\
 6609.1104 & 2.559 & -2.692 & -2.635 \\
 6810.2627 & 4.607 & -1.000 & -0.959 \\
 6828.5913 & 4.638 & -0.820 & -0.810 \\
 6841.3389 & 4.607 & -0.710 & -0.645 \\
 6843.6558 & 4.548 & -0.830 & -0.817 \\
 6858.1499 & 4.607 & -0.940 & -0.931 \\
 8327.0674 & 2.200 & -1.525 & -1.575 \\
 8688.6426 & 2.170 & -1.212 & -1.236 \\
\enddata
\label{tab:table2}
\end{deluxetable}

\clearpage
\begin{deluxetable}{cccccccccccc}
\tabletypesize{\scriptsize}
\tablecolumns{12}
\tablewidth{0pc}
\tablecaption{Derived stellar parameters with corresponding uncertainties and [Fe/H] error components for the solar-similar primaries.} 
\tablehead{
 \colhead{Name} & 
 \colhead{\textsl{T}$_{eff}$} & 
 \colhead{$\sigma$} &
 \colhead{log \textsl{g}} & 
 \colhead{$\sigma$} &
 \colhead{[Fe/H]} & 
 \colhead{$\sigma_{l}$} &
 \colhead{$\sigma_{t}$} &
 \colhead{$\sigma_{g}$} &
 \colhead{$\sigma$} &
 \colhead{$\xi$} &
 \colhead{$\eta$} \\
 \colhead{} & 
 \colhead{(K)} & 
 \colhead{} &
 \colhead{(cgs)} & 
 \colhead{} &
 \colhead{} & 
 \colhead{} &
 \colhead{} &
 \colhead{} &
 \colhead{} &
 \colhead{(km s$^{-1}$)} &
 \colhead{(km s$^{-1}$)}
}
\startdata
Sun        & 5777\tablenotemark{a} & \nodata & 4.44\tablenotemark{a} & \nodata &  0.00\tablenotemark{a,b} & \nodata & \nodata & \nodata & \nodata & 1.24 & 1.90 \\
HIP 12114  & 4867 & 119 & 4.64 & 0.04 & -0.12 & 0.01 & 0.04 & 0.02 & 0.05 & 0.84 & 1.44 \\
HIP 26907  & 5054 & 122 & 4.60 & 0.10 & +0.03 & 0.01 & 0.06 & 0.01 & 0.06 & 0.88 & 1.60 \\
HIP 32423  & 4730 & 117 & 4.66 & 0.03 & -0.23 & 0.01 & 0.02 & 0.01 & 0.03 & 0.69 & 0.81 \\
HIP 40035  & 6262 & 102 & 4.30 & 0.10 & -0.02 & 0.02 & 0.06 & 0.02 & 0.07 & 1.69 & 6.52 \\
HIP 102040 & 5737 & 111 & 4.49 & 0.22 & -0.14 & 0.01 & 0.06 & 0.01 & 0.06 & 1.16 & 2.10 \\
\enddata
\label{tab:table3}
\tablenotetext{a}{Fixed.}
\tablenotetext{b}{$log\ \epsilon$(Fe) = 7.45.}
\end{deluxetable}

\clearpage
\begin{deluxetable}{cccccccccc}
\tabletypesize{\scriptsize}
\tablecolumns{10}
\tablewidth{0pc}
\tablecaption{Stellar parameters for the 30 test sample stars from \citet{carlos04}, \citet{vf05}, and this study.}
\tablehead{
 \colhead{} &
 \multicolumn{3}{c}{\citet{carlos04}} &
 \multicolumn{3}{c}{\citet{vf05}} &
 \multicolumn{3}{c}{This Study} \\
 \colhead{Name} & 
 \colhead{\textsl{T}$_{eff}$} & 
 \colhead{log \textsl{g}} & 
 \colhead{[Fe/H]} &
 \colhead{\textsl{T}$_{eff}$} & 
 \colhead{log \textsl{g}} & 
 \colhead{[Fe/H]} &
 \colhead{\textsl{T}$_{eff}$} & 
 \colhead{log \textsl{g}} & 
 \colhead{[Fe/H]} \\
 \colhead{(HIP)} & 
 \colhead{(K)} & 
 \colhead{(cgs)} & 
 \colhead{} &
 \colhead{(K)} & 
 \colhead{(cgs)} & 
 \colhead{} &
 \colhead{(K)} & 
 \colhead{(cgs)} & 
 \colhead{} 
}
\startdata
  544 & 5353 & 4.55 &  0.03 & 5577 & 4.58 &  0.11 & 5360 & 4.54 &  0.00 \\
 3093 & 5117 & 4.58 &  0.13 & 5221 & 4.45 &  0.16 & 5144 & 4.56 &  0.15 \\
 3765 & 4980 & 4.65 & -0.25 & 4944 & 4.51 & -0.27 & 4881 & 4.65 & -0.30 \\
 3821 & 5801 & 4.47 & -0.40 & 5941 & 4.44 & -0.25 & 5710 & 4.30 & -0.41 \\
 7513 & 6100 & 4.17 &  0.02 & 6213 & 4.25 &  0.15 & 6089 & 4.12 &  0.03 \\
 7981 & 5138 & 4.60 & -0.04 & 5181 & 4.53 & -0.04 & 5088 & 4.59 & -0.06 \\
 8362 & 5257 & 4.58 & -0.04 & 5327 & 4.54 &  0.03 & 5215 & 4.57 &  0.00 \\
12777 & 6210 & 4.35 & -0.08 & 6344 & 4.42 &  0.06 & 6162 & 4.28 & -0.06 \\
14632 & 5877 & 4.27 &  0.01 & 6032 & 4.31 &  0.16 & 5853 & 4.20 &  0.01 \\
15457 & 5564 & 4.52 & -0.11 & 5742 & 4.49 &  0.12 & 5553 & 4.50 & -0.05 \\
16537 & 5052 & 4.62 & -0.06 & 5146 & 4.57 & -0.03 & 4988 & 4.62 & -0.05 \\
16852 & 5914 & 4.12 & -0.17 & 6038 & 4.21 & -0.02 & 5824 & 4.03 & -0.21 \\
17420 & 4801 & 4.63 & -0.02 & 4991 & 4.59 & -0.06 & 4869 & 4.63 & -0.04 \\
19849 & 5164 & 4.61 & -0.17 & 5151 & 4.57 & -0.28 & 5046 & 4.61 & -0.34 \\
22449 & 6424 & 4.34 &  0.00 & 6424 & 4.29 &  0.03 & 6300 & 4.27 & -0.05 \\
23311 & 4641 & 4.63 &  0.26 & 4827 & 4.69 &  0.33 & 4746 & 4.61 &  0.34 \\
24813 & 5781 & 4.33 & -0.01 & 5911 & 4.37 &  0.12 & 5737 & 4.24 & -0.01 \\
26779 & 5150 & 4.58 &  0.14 & 5351 & 4.60 &  0.19 & 5145 & 4.58 &  0.09 \\
37349 & 4889 & 4.62 &  0.04 & 4964 & 4.71 &  0.07 & 5012 & 4.63 &  0.09 \\
27913 & 5820 & 4.49 & -0.17 & 5882 & 4.34 & -0.01 & 5818 & 4.47 & -0.15 \\
37279 & 6677 & 4.08 &  0.03 & 6543 & 3.99 &  0.00 & 6482 & 4.00 & -0.13 \\
40693 & 5331 & 4.57 & -0.12 & 5361 & 4.46 & -0.06 & 5340 & 4.56 & -0.05 \\
43587 & 5063 & 4.56 &  0.34 & 5235 & 4.45 &  0.31 & 5173 & 4.53 &  0.34 \\
51459 & 6057 & 4.43 & -0.17 & 6126 & 4.34 & -0.07 & 5968 & 4.35 & -0.23 \\
53721 & 5751 & 4.35 & -0.10 & 5882 & 4.38 &  0.04 & 5740 & 4.26 & -0.07 \\
56997 & 5402 & 4.57 & -0.16 & 5488 & 4.43 & -0.03 & 5409 & 4.56 & -0.10 \\
57757 & 6076 & 4.14 &  0.08 & 6161 & 4.22 &  0.18 & 6252 & 4.18 &  0.19 \\
58576 & 5361 & 4.47 &  0.14 & 5565 & 4.56 &  0.27 & 5428 & 4.45 &  0.21 \\
61317 & 5743 & 4.47 & -0.35 & 5930 & 4.44 & -0.16 & 5748 & 4.37 & -0.30 \\
64394 & 5910 & 4.44 & -0.05 & 6075 & 4.57 &  0.07 & 5956 & 4.43 & -0.00 \\
\enddata
\label{tab:table4}
\end{deluxetable}

\clearpage
\begin{deluxetable}{ccccc}
\tabletypesize{\scriptsize}
\tablecolumns{5}
\tablewidth{0pc}
\tablecaption{Species, central wavelengths ($\lambda$), lower level excitation potential ($\chi$), astrophysical log \textsl{gf}s, and damping source for the lines used in the modified analysis of the M dwarfs. The damping types are ``barklem,'' which refers to data from \citet{barklem00} and ``uns\"{o}ld,'' which refers to the approximation of \citet{unsold55} enhanced by a factor of 2.5.}
\tablehead{
 \colhead{Species} & 
 \colhead{$\lambda$} &
 \colhead{$\chi$} &
 \colhead{log \textsl{gf}} &
 \colhead{Damping} \\
 \colhead{} & 
 \colhead{(\AA)} & 
 \colhead{(eV)} 
}
\startdata
 Fe \textsc{i}\tablenotemark{a}  & 8327.067 & 2.20 & -1.575 & barklem \\
 Ti \textsc{i}  & 8364.237 & 0.84 & -1.684 & uns\"{o}ld \\
 Ti \textsc{i}  & 8377.861 & 0.83 & -1.521 & uns\"{o}ld \\
 Ti \textsc{i}  & 8382.530 & 0.82 & -1.549 & uns\"{o}ld \\
 Ti \textsc{i}  & 8382.780 & 0.81 & -1.652 & uns\"{o}ld \\
 Fe \textsc{i}  & 8387.772 & 2.18 & -1.562 & barklem \\
 Ti \textsc{i}  & 8396.898 & 0.81 & -1.646 & uns\"{o}ld \\
 Ti \textsc{i}  & 8412.358 & 0.82 & -1.376 & barklem \\
 Ti \textsc{i}  & 8426.506 & 0.83 & -1.136 & barklem \\
 Fe \textsc{i}  & 8661.897 & 2.22 & -1.537 & barklem \\
 Ca \textsc{ii} & 8662.141 & 1.69 & -0.716 & barklem \\
 Fe \textsc{i}\tablenotemark{b}  & 8674.746 & 2.83 & -1.846 & barklem \\
 Ti \textsc{i} \tablenotemark{b} & 8675.372 & 1.07 & -1.465 & barklem \\
 Ti \textsc{i}\tablenotemark{b}  & 8682.980 & 1.05 & -1.762 & barklem \\
 Fe \textsc{i}\tablenotemark{a,b}  & 8688.643 & 2.17 & -1.236 & barklem \\
 Ti \textsc{i}\tablenotemark{b} & 8692.331 & 1.05 & -2.098 & barklem \\
\enddata
\label{tab:table5}
\tablenotetext{a}{Also used in the analysis of the solar-similar primaries.}
\tablenotetext{b}{Used in the original V98 analysis.}
\end{deluxetable}

\clearpage
\begin{deluxetable}{ccccccc}
\tabletypesize{\scriptsize}
\tablecolumns{7}
\tablewidth{0pc}
\tablecaption{Derived stellar parameters for the M dwarf secondaries using the revised technique. Deviations in [M/H] ($\equiv$ [Fe/H]) from the value of the corresponding primary are also given. Adopted uncertainties are 48 K, 0.10 dex, 0.12 dex, 0.15 km s$^{-1}$, and 0.20 km s$^{-1}$ for {T}$_{eff}$, log \textsl{g}, [M/H], $\xi$, and $\eta$ respectively.}
\tablehead{
 \colhead{Name} & 
 \colhead{\textsl{T}$_{eff}$} & 
 \colhead{log \textsl{g}} & 
 \colhead{[M/H]} & 
 \colhead{$\xi$} &
 \colhead{$\eta$} &
 \colhead{$\Delta$([M/H])\tablenotemark{a}} \\
 \colhead{} & 
 \colhead{(K)} & 
 \colhead{(cgs)} & 
 \colhead{} &
 \colhead{(km s$^{-1}$)} &
 \colhead{(km s$^{-1}$)} &
 \colhead{}
}
\startdata
HIP 32423B  & 3722 & 4.82 & -0.31 & 0.94 & 0.73 & -0.08 \\
HIP 40035B  & 3659 & 4.72 & -0.12 & 0.91 & 1.60 & -0.10 \\
HIP 102040B & 3556 & 4.76 & -0.21 & 0.83 & 1.41 & -0.07 \\
HIP 26907B  & 3531 & 4.82 & -0.13 & 0.91 & 0.82 & -0.16 \\
HIP 12114B  & 3444 & 4.98 & -0.09 & 0.85 & 0.62 & +0.03 \\
\enddata
\label{tab:table6}
\tablenotetext{a}{Deviation with respect to the value of the primary given in Table ~\ref{tab:table3}.}
\end{deluxetable}

\clearpage

\clearpage

\clearpage

\clearpage

\clearpage

\clearpage

\clearpage

\clearpage


\newpage
\begin{thebibliography}{}

\bibitem[Allard et al.(2000)]{allard00}Allard, F., Hauschildt, P. H., \& Schwenke, D. 2004, \apj, 540, 1005

\bibitem[Allen et al.(2000)]{allen00}Allen, C., Poveda, A., \& Herrera, M. A. 2000, \aap, 356, 529

\bibitem[Allende Prieto et al.(2004)]{carlos04}Allende Prieto, C., Barklem, P. S., Lambert, D. L., \& Cunha, K. 2004, \aap, 420, 183

\bibitem[Anders \& Grevesse(1989)]{anders89}Anders, E., \& Grevesse, N. 1989, Geochim. Cosmochim. Acta, 53, 197

\bibitem[Asplund et al.(1997)]{asplund97}Asplund, M., Gustofsson, B., Kiselman, D., \& Eriksson, K. 1997, \aap, 318, 521

\bibitem[Asplund et al.(2005)]{asplund05}Asplund, M., Grevesse, N., \& Sauval, A. J. 2005, in ASP Conf. Ser. 336, Cosmic Abundances as Records of Stellar Evolution and Nucleosynthesis in honor of David L. Lambert, ed. A. G. Barnes III \& F. N. Bash (San Francisco: ASP), 25

\bibitem[Baraffe et al.(1998)]{baraffe98}Baraffe, I., Chabrier, G., Allard, F., \& Hauschildt, P.H. 1998, \aap, 337, 403

\bibitem[Barklem et al.(2000)]{barklem00}Barklem, P. S., Piskunov, N., \& O'Mara, B. J. 2000, \aaps, 142, 467

\bibitem[Benedict et al.(2002)]{benedict02}Benedict, G. F., \& et al. 2002, \apjl, 581, 115

\bibitem[Berger et al.(2006)]{berger06}Berger, D. H., \& et al. 2006, \apj, in press

\bibitem[Bertelli et al.(1994)]{bertelli94}Bertelli, G., Bressan, A., Chiosi, C., Fagotto, F., \& Nasi, E. 1994, \aaps, 106, 275

\bibitem[Bertone et al.(2004)]{bertone04}Bertone,E., Buzzoni, A., Chávez, M., \& Rodríguez-Merino, L. H. 2004, \aj, 128, 829 

\bibitem[Bonfils et al.(2005)]{bonfils05}Bonfils, X., \& et al. 2005, \aap, 443, 15

\bibitem[Butler et al.(2004)]{butler04}Butler, R. P., Vogt, S. S., Marcy, G. W., Fischer, D. A., Wright, J. T., Henry, G. W., Laughlin, G., \& Lissauer, J. J. 2004, \apjl, 617, 580

\bibitem[Cayrel de Strobel et al.(2001)]{cayrel01}Cayrel de Strobel, G., Soubiran, C., \& Ralite, N. 2001, \aap, 373, 159

\bibitem[Chase et al.(1985)]{janaf85}Chase Jr., M. W., Davies, C. A., Downey Jr., J. R., Frurip, D. J., McDonald, R. A, \& Syverud, A. N. 1985, JANF Thermochemical Tables, 3rd Ed., J. Phys. Chem. Ref. Data 14, Suppl. 1

\bibitem[Creevey et al.(2005)]{creevey05}Creevey, O. L., \& et al. 2005, \apj, 625, L127

\bibitem[Cutri et al.(2003)]{cutri03}Cutri, R. M., \& et al. 2003 The 2MASS All-Sky Catalog of Point Sources (Pasadena: IPAC/California Inst. Technology)

\bibitem[Dawson \& De Robertis (2004)]{dawson04}Dawson, P. C., \& De Robertis, M. M. 2004, 127, 2909

\bibitem[Delfosse et al.(1998)]{delfosse98}Delfosse, X., Forveille, T., Mayor, M., Perrier, C., Naef, D., \& Queloz, D. 1998, \aap, 338, 67

\bibitem[Delfosse et al.(2000)]{delfosse00}Delfosse, X., Forveille, T., S\'{e}gransan, D., Beuzit, J. -L., Udry, S., Perrier, C., \& Mayor, M. 2000, \aap, 364, 217

\bibitem[Desidera et al.(2004)]{desidera04}Desidera, S., \& et al. 2004, \aap, 420, 683

\bibitem[Dubois \& Gole(1977)]{dubois77}Dubois, L. H., \& Gole, J. L. 1977, J. Chem. Phys. 66, 779

\bibitem[ESA (1997)]{hipp}ESA 1997, The Hipparcos and Tycho Catalogues, ESA SP-1200

\bibitem[Fischer \& Valenti(2005)]{fv05}Fischer, D. A., \& Valenti, J. A. 2005, \apj, 622, 1102

\bibitem[Gustafsson et al.(1975)]{gus75}Gustafsson, B., Bell, R. A., Eriksson, K., \& Nordlund, \AA. 1975, \aap, 42, 407

\bibitem[Hauschildt et al.(1999)]{haus99}Hauschildt, P. H., Allard, F., \& Baron, E. 1999, \apj, 512, 377

\bibitem[Henry (1998)]{henry98}Henry, T. J. 1998, in ASP Conf. Ser. 134, Brown Dwarfs and Extrasolar Planets, ed. R. Rebolo, E. L. Mart\'{i}n, \& M. R. Zapatero Osorio (San Francisco: ASP), 28

\bibitem[Henry et al.(1998)]{henry99}Henry, T. J., \& et al. 1999, \apj, 512, 864

\bibitem[Holmberg et al.(2006)]{holmberg06}Holmberg, J., Flynn, C., \& Portinari, L. 2006, \mnras, 367, 449

\bibitem[Huber \& Herzberg(1979)]{huber79}Huber, K. P., \& Herzberg, G. 1979, Molecular Spectra and Molecular Structure, Vol. 4, Constants of Diatomic Molecules (New York: Van Nostrand Rheinhold)

\bibitem[J\o rgensen (1994)]{jor94}J\o rgensen, U. G. 1994, \aap, 284, 179

\bibitem[Kroupa \& Tout (1997)]{kroupa97}Kroupa, P., \& Tout, C. A. 1997, \mnras, 287, 402

\bibitem[Kupka et al.(1999)]{kupka99}Kupka, F., Pisknunov, N. E., Ryabchikova, T. A., Stempels, H. C., \& Weiss, W. W. 1999, \aaps, 138, 119

\bibitem[Kurucz et al.(1984)]{kurucz84}Kurucz, R. L., Furenlid, I., Brault, J., \& Testerman, L. 1984, National Solar Observatory Atlas (Sunspot, New Mexico: NSO)

\bibitem[Kurucz (1993)]{kurucz93}Kurucz, R. L. 1993, ATLAS9 Stellar Atmosphere Programs and 2 km/sec grid, Kurucz CD-ROM No. 13

\bibitem[Kurucz (1999)]{kurucz99}Kurucz, R. L. 1999, TiO linelist from Schwenke, Kurucz CD-ROM No. 24

\bibitem[Ku\v{c}inskas et al.(2005)]{kucinskas05}Ku\v{c}inskas, A., Hauschildt, P. H., Ludwig, H.-G., Brott, I., Vansevi\v{c}ius, V, Lindegren, L, Tanab\'{e}, T, \& Allard, F. 2005, \aap, 442, 281.

\bibitem[Lane et al.(2001)]{lane01}Lane, B. F., Boden, A. F., \& Kulkarni, S. R. 2001, \apj, 551, L81

\bibitem[Leggett et al. (1996)]{leggett96}Leggett, S. K., Allard, F., Berriman, G., Dahn, C. C., \& Hauschildt, P. H. 1996, \apjs, 104, 117

\bibitem[L\'{o}pez-Morales \& Ribas(2005)]{lopez05}L\'{o}pez-Morales, M., \& Ribas, I. 2005, \apj, 631, 1120

\bibitem[Maceroni \& Montalb\'{a}n(2004)]{maceroni04}Maceroni, C., \& Montalb\'{a}n, J. 2004, \aap, 426, 577

\bibitem[Marcy et al. (1998)]{marcy98}Marcy, G. W., Butler, R. P., Vogt, S. S., Fischer, D. A., \& Lissauer, J. J. 1998, \apjl, 505, 147

\bibitem[Marcy et al. (2001)]{marcy01}Marcy, G. W., Butler, R. P., Fischer, D. A., Vogt, S. S., Lissauer, J. J., \& Rivera, E. J. 2001, \apj, 556, 296

\bibitem[Marquardt(1963)]{mar63}Marquardt, D. W. 1963, J. Soc. Ind. Appl. Math. 11, 431

\bibitem[Maxted et al.(2004)]{maxted04}Maxted, P. F. L., \& et al. 2004, \mnras, 355, 1143

\bibitem[Metcalfe et al.(1996)]{metcalfe96}Metcalfe, T. S., Mathieu, R. D., Latham, D. W., \& Torres, G. 1996, \apj, 456, 356

\bibitem[Naulin et al.(1997)]{naulin97}Naulin, C., Hedgecock, I. M.,\& Costes, M. 1997, Chem. Phys. L., 266, 335

\bibitem[Perryman et al.(1997)]{perryman97}Perryman, M. A. C. \& et al. 1997, \aap, 323, 49

\bibitem[Piskunov (1992)]{piskunov92}Piskunov, N. E. 1992 in Stellar Magnetism, Proceedings of international meeting on the problem "Physics and evolution of stars", ed. Yu.V. Glagolevskij \& I.I. Romanyuk (Saint Petersburg: NAUKA), 92

\bibitem[Piskunov et al.(1995)]{piskunov95}Piskunov, N. E., Kupka, F., Ryabchikova, T. A., Weiss, W. W., \& Jeffery, C. S. 1995, \aaps, 112, 525 

\bibitem[Plez (1998)]{plez98}Plez, B. 1998, \aap, 337, 495

\bibitem[Poveda et al. (1994)]{poveda94}Poveda, A., Herrera, M. A., Allen, C., Cordero, G., Lavalley, C. 1994, RMxAA, 28, 43

\bibitem[Press et al.(1986)]{press86}Press, W. H., Flannery, B. P., Teukolsky, S. A., \& Vetterling W. T. 1986, Numerical Recipes (Cambridge: Cambridge University Press)

\bibitem[Ramirez \& Mel\'{e}ndez(2004)]{ramirez04}Ramirez, I. \& Mel\'{e}ndez, J 2004, \apj, 609, 417

\bibitem[Ramirez et al.(2006)]{ramirez06}Ramirez, I., Allende Prieto, C., \& Lambert, D. L. 2006, in preparation

\bibitem[Ramirez \& Mel\'{e}ndez(2005)]{ramirez05}Ramirez, I. \& Mel\'{e}ndez, J 2005, \apj, 626, 465

\bibitem[Reddy et al.(2003)]{reddy03}Reddy, B. E., Tomkin, J., Lambert, D. L., \& Allende Prieto, C. 2003, \mnras, 340, 304

\bibitem[Reid \& Hawley (2005)]{reidhawley04}Reid, I. N., \& Hawley, S. L. 2005, New Light on Dark Stars: Red Dwarfs, Low-Mass Stars, Brown Dwarfs (2nd ed.; London: Springer)

\bibitem[Ribas (2003)]{ribas03}Ribas, I. 2003, \aap, 398, 239

\bibitem[Rivera et al. (2005)]{rivera05}Rivera, E. J., \& et al 2005, \apjl, 634, 625

\bibitem[Santos et al.(2005)]{santos05}Santos, N. C., Israelian, G., Mayor, M., Bento, J. P., Almeida, P. C., Sousa, S. G., \& Ecuvillon, A. 2005, \aap, 437, 1127

\bibitem[Sauval \& Tatum(1984)]{sauval84}Sauval, A. J., \& Tatum, J. B. 1984, \apjs, 56, 193

\bibitem[Schweitzer et al.(1996)]{schweitzer96}Schweitzer, A., Hauschildt, P., Allard, F., \& Basri, G. 1996, \mnras, 283, 821

\bibitem[S\'{e}gransan et al.(2003)]{seg03}S\'{e}gransan, D., Kervella, P., Forveille, T., \& Queloz, D. 2003b, \aap, 397, L5

\bibitem[Sekiguchi \& Fukugita (2000)]{sekiguchi00}Sekiguchi, M., \& Fukugita, M. 2000, \aj, 120, 1072

\bibitem[Siess et al.(2000)]{siess00}Siess, L., Dufour, E., \& Forestini, M. 2000, \aap, 358, 593

\bibitem[Sneden (1973)]{sneden73}Sneden, C. 1973, Ph.D. Thesis, University of Texas

\bibitem[Torres \& Ribas(2002)]{torres02}Torres, G., \& Ribas, I. 2002, \apj, 567, 1140

\bibitem[Tull et al. (1995)]{tull95}Tull, R. G., MacQueen, P. J., Sneden, C., \& Lambert, D. L. 1995, \pasp, 107, 251

\bibitem[Valenti \& Fischer (2005)]{vf05}Valenti, J. A., \& Fischer, D. A. 2005, \apjs, 159, 141

\bibitem[Uns\"{o}ld (1955)]{unsold55}Uns\"{o}ld, A. 1955, Physik der Sternatmosph\"{a}ren. Springer-Verlag, Berlin

\bibitem[Valenti, Piskunov, \& Johns-Krull (1998)]{v98}Valenti, J., Piskunov, N., \& Johns-Krull, C. M. 1998, \apj, 498, 851, V98

\bibitem[Woolf \& Wallerstein (2006)]{woolf06}Woolf, V. M., \& Wallerstein, G. 2006, \pasp, 118, 218


\end{thebibliography}
\end{document}